\documentclass[Times,12pt,oneside,openany,print,index]{report}
\usepackage[a4paper,width=150mm,top=25mm,bottom=25mm]{geometry}
\usepackage[english]{babel}
\usepackage[utf8]{inputenc}
\usepackage{csquotes} 
\usepackage{amsmath} 
\usepackage{graphicx} 
\graphicspath{/images} 
\usepackage{caption} 
\usepackage{array} 

\usepackage[nottoc]{tocbibind} 

\usepackage[normalem]{ulem} 

\usepackage{hyperref} 
\hypersetup{
    colorlinks=true,
    linkcolor=black,
    filecolor=magenta,      
    urlcolor=black,
    citecolor=black,
}
\usepackage[backend=biber,style=ieee,sorting=ynt]{biblatex} 
\let\cleardoublepage=\clearpage 

\begin{document}

\thispagestyle{empty} 
\begin{titlepage}
\renewcommand*{\thepage}{Title} 

    \begin{center} 
        \vspace*{3cm} 
        
        {\fontsize{16pt}{22pt}\selectfont{Sizing Strategies for Algorithmic Trading in Volatile Markets: A Study of Backtesting and Risk Mitigation Analysis}
        } 
        
        \vspace{1.0cm}
        
        \text{by}
        
        \vspace{0.5cm}
        
        	S. M. Masrur Ahmed\\
	        Brac University, Dhaka, Bangladesh\\
	        s.m.masrur.ahmed@g.bracu.ac.bd
    \end{center}

\end{titlepage} 
\cleardoublepage

\pagenumbering{roman} 

\phantomsection
\addcontentsline{toc}{chapter}{Abstract}
\section*{Abstract}
Backtest is a way of financial risk evaluation which helps to analyze how our trading algorithm would work in markets with past time frame. The high volatility situation has always been a critical situation which creates challenges for algorithmic traders. The paper investigates different models of sizing in financial trading and backtest to high volatility situations to understand how sizing models can lower the models of \emph{VaR} during crisis events. Hence it tries to show that how crisis events with high volatility can be controlled using short and long positional size. The paper also investigates stocks with AR, ARIMA, LSTM, GARCH with ETF data.

\vspace{1cm}
\textbf{Keywords:} Quantative Finance; Algorithmic Trading; Positional Size; Value-at-risk; Kalman-filter; BollingerBands;Stock Technical Indicator; Backtest; Bayesian Analysis; Risk model;
\pagebreak

\renewcommand{\contentsname}{Table of Contents} 
\cleardoublepage
\phantomsection
\addcontentsline{toc}{chapter}{Table of Contents} 
\tableofcontents

\pagenumbering{arabic} 

\chapter{Introduction}
\section{Introduction}
\section{Financial Bubbles and Manias } 
Stock market bubbles are a well-known phenomenon worldwide. In the financial world, a bubble occurs when the price of a bond, asset, or real estate significantly exceeds its intrinsic value, often driven by irrational exuberance. When such a bubble bursts, it can have severe consequences, leading to frantic selling by investors. In the world of finance, a bubble refers to a period when the prices of stocks, bonds, or real estate irrationally surpass the assets' intrinsic values. It is known since \begin{math}{17^{th}}\end{math} century’s "tulip crazy" in dutch to America’s outbreak of the subprime crisis in 2007. The stock price bubbles has always put algorithmic trades in challenge on how to deal with it while minimizing the VaR. Thus, it became more challenging to research on minimizing volatility of portfolios. For a financial institution like bank, it is very essential to take size and risk into consideration to fix the necessary amount of capital for assets. This is also very important for stock markets as well.

\section{Aims and Objectives }
The primary aim of this study is to develop sizing models for financial trading and backtest to high volatility situations to understand how sizing models can lower the VaR during crisis events on stock and ETF market. 
\section{History of Financial Trading Market}\
In 1791, United states introduced the first stock exchange. It was in the Philadelphia which was their leading city for foreign and domestic trade. New York's stock exchange was established 1792, when around 24 brokers and merchants agreed to get a commission for being a agents to give support to one another for negotiations. Under a 68, Wall Street tree, they performed their trading. Trading during day time has been for centuries. Today we use electronics device or internet to trade but buying and selling securities goes back to \begin{math}{19^{th}}\end{math} century. People could make trade on small scale businesses around the major cities which was known as bucket shops. The traders use to contribute their money into the common bucket. Stocks or commodities with leverage were bought using that pool of money. The process was quite different than now. There were clerks who wrote prices on a chalkboard while another one read the tickers. Winning or loosing money solely depended on the honesty of the shop or business operators. In 1929 those shops of small businesses died with the crash of stock market and got regulated out of stock market.
\\ \\
In 1930, The modern stock market was reopened with new laws for strict regulation. From that with many crashes and new laws, today's major markets were established which  are \emph{NYSE, NASDQ, PSE, PHLX, CBOT} etc. Apart from America, Germany's \emph{DAX}, France's \emph{CAC}, Hong Kong's \emph{Hang Sheng},Japan's \emph{Nikkei},London's \emph{FTSE}, Dhaka's \emph{DSE} and many more. A commission or exchange fee needs to be paid for any trade of stocks. 

\section{Backtesting Framework}
Backtesting framework is mostly used by the banks for the market's risk measurement. It is also employed by algorithmic traders and strategists. It actually illustrate periodic comparison among VaR and Returns which can be profit or loss. They compare trading outcomes with VaR to check how the model is performing and putting us to risk. It also counts how many times VaR exceeded the trading outcome on historical time frame. The risk analysis of backtest shows an amount that can be lost over a period using a specified confidence level.In VaR Model, trading time close value is given as the input. It analyzes on the change in the value of portfolio for due to rate and price changes. This also provides judgements on the sensitivity of the portfolio. This concept actually makes the backtesting a bit complicated. It is argued that actual trading outcomes can not be compared with VaR analysis because actual outcome may get corrupted by the structural changes of portfolio during hold position. 
\\ \\ 
Backtesting a method for integral calculation of statistical test to measure VaR which takes place in hypothetical changes affecting the portfolio for the stock market. To face the situation where risk analysis fails to capture the trading volatility, actual losses and profits can be used as a parameter for backtesting to calculate with integrity. Backtesting is sometimes referred to as a parametric test for $\alpha$, denoting the probability of coverage, defining the quantile VaR measure. More formally, let $Y_t$ be the portfolio returns which stands for profit or losses for time series and assumes that at $t-1$ time information set for agent is given by a $d_w$ dimension of random vector $w_{t-1}$ which can contain past values for $Y_t$, that is $w_{t-1}=\{Y_x,Z_x^{\prime}\}_{x=t-h^{\prime}}^{t-1}$, $h<\infty$. Here the transpose matrix for $Z$ is denoted as $Z^{\prime}$. If the distribution of $Y_t$ is continuous, we can say $v_\alpha(w_{t-1},\theta)$ is a correctly specified $\alpha^{th}$ VaR.
\begin{equation}
    p(Y_t \leq v_\alpha(w_{t-1},\theta_{0}) | F_{t-1}) = \alpha \begin{text}{ where }\end{text} \alpha \in (0,1)\: \forall\;t\in\;Z
\end{equation}
\\ \\

The frequency and nature of backtesting are crucial considerations. The simplest way is to calculate the number of times when the trading outcomes were not covered by risk analysis. 
\\ \\
The market capital amount for backtesting is also important. It changes the amount of loss or profit. In this paper's simulations, a capital amount of \$10 million was employed It gives a reasonable fluidity to the simulation for backtesting.
\\ \\
The paper used the bactesting for high volatile market situation during crisis events. Even when the paper is being written we are in a crisis moment of corona virus pandemic. This paper specifically focuses on backtesting in high-volatility market situations, particularly during crisis events. The study places emphasis on sizing values for both short and long stocks and examines their impact on reducing risk metrics while achieving satisfactory returns. The paper will discuss the size values for short and long stocks and backtest it to see how it lowers the risk metrics for a considerable returns.

\chapter{Related Work}
The optimal position sizing ideas go back to the year 1956 where it was illustrated that the application of Information Theory on betting and the paper proved that the information transfer rate is equal to the rate of exponential growth over a channel of gambler's capital \cite{criterion35kelly}.However, the original theory of the Kelly criterion cannot be applicable in finance because the output of the theory is not Bernoulli distributed, unlike in gambling games. In finance the growth rate can be maximized with only logarithmic utility. The idea of maximizing utility through a logarithmic function which delivers an optimal rate of growth in an economic context \cite{latane1959criteria}. A well known mathematician, Edward Thorp from the United States adopted the Kelly criterion formula for portfolio selection \cite{thorp1969optimal,thorp1975portfolio,thorp1980kelly,maclean2011kelly}. There were other works which were done by other authors who elaborated the concept further and proved implementable solutions \cite{browne1999risk,hakansson1995capital,maclean2011kelly,maclean2011time,wilcox2003harry}. 
\\ \\
"Calculating VaR for a portfolio is categorized into three \cite{pilar9comprehensive}. They are: \begin{itemize}
  \item Historical Simulation or Non-Parametric process
  \item Variance \& Co-variance approach or The Parametric process
  \item Monte-Carlo simulation or Semi-Parametric process
\end{itemize}
The paper also noted that these models standardize the shortcomings that have created the development of new proposals. The paper also discussed that the Variance-Covariance method has a major drawback for VaR estimation as the model assumes normal distribution for returns on investment. This model also has a relation to conditional volatility. The paper also suggests that parametric methods have grown much to get away from drawbacks. The paper also explains that the historical result simulation has better density estimation.
\\
\\
The Filtered Historical Simulation, which falls under Semi-Parametric Simulation, was proposed by \cite{barone1999var} which was applied by Sommacampagna \cite{sommacampagna2002estimating}, where the author estimated VaR, Conditional Autoregressive VaR, and different approaches for Extreme Value Theory.
\\ \\
There are several models of return distribution that can be compared with the predictive ability of Value at Risk. Doric and Doric used the means of backtest models for the sample and found symmetric behavior of returns in the case of stock data \cite{doric2016return}. They found that Student’s T-Distribution and NIG distribution are suitable for $\alpha$ values.They also expressed that parametric models which are unconditional assume that returns are independently and identically distributed, where density is,
\begin{equation}
    f_z(z)=\frac{1}{\sigma}f_{k^*}(\frac{z-\mu}{\sigma})
\end{equation}
where $f_k$ is the function of density for the distribution of $k_t$ and $f_{k^*}$is the standardized distribution of $k_t$ of density function. The parameters $\mu$ and $\sigma$ are respectively trend and volatility of $k_t$. The VaR for a long trading position is given by,
\begin{equation}
    VaR_{l} = \mu + k_\alpha^*\sigma
\end{equation}
and for a short position it is, 
\begin{equation}
    VaR_{s} = \mu + k_{1-\alpha}^*\sigma
\end{equation}
where, $k_\alpha^*$ denotes the $\alpha$-quantile of $f_k^*$, and $\alpha= 0.05$.

\chapter{Backtesting Methodologies}
Backtesting is a process in which analytical methods or trading strategies are applied to historical data to see how accurate the portfolio is working or measuring VaR. In this section, few backtesting processes will be discussed. 
\newtheorem{theorem}{Theorem}
\section{Bernoulli Trials (Kelly, 1956)}
This test was introduced by Kelly using a logarithmic utility function for gambling and information theory \cite{criterion35kelly}. Bernoulli trials, also known as Binary Channel, demonstrated that the logarithmic utility function can maximize the growth rate. However, it can lead to myopic decision-making when used for short-term periods \cite{maclean2011kelly}.\\

Assuming in a gamble, the winning probability is $\frac{1}{2}<p\leq1$ and the outcome is 1 and probability of loosing is $q=1-p$ with $-1$ output, starting with the capital $C_0$. The capital after the $n^th$ trial, gambling fraction $g$ of the initial capital (in \%),is given by,
\begin{equation}
    C_n = C_0(1+g)^m(1-g)^{-m+n}
\end{equation}
The growth for each trial which is exponential from equation $3.1$,
\begin{equation}
    G_n(g)=log(\frac{C_n}{C0})^{\frac{1}{n}}=log[(1+g)^{\frac{m}{n}}(1-g)^\frac{n-m}{n}]\\=(\frac{m}{n})log(1+g)+(\frac{n-m}{n})log(1-g)
\end{equation}
\\
Also, the coefficient for the anticipated growth rate is,
\begin{equation}
    E[G_n(g)]=\gamma(g)=p.log(1+g)+q.log(1-g)=E[log(C)]
\end{equation}
if we maximize $\gamma(g)$, we get,
\begin{align}\gamma^\prime(g)=(\frac{p}{1+g})-(\frac{q}{1-g})=[\frac{p-q-g}{(1-g^2)}]\\
    \longleftrightarrow g = g^* = p - q , p\geq q >0  \nonumber
\end{align}
After obtaining the second derivative of $f$, it becomes evident that $g = g^*$ represents the maximum and is unique as well,
\begin{equation}
    \gamma(g^*) = p.log(p) + q.log(q) + log(2) > 0 
\end{equation}
\begin{equation}
    \gamma^{\prime\prime}(g)= - [\frac{p}{(1+g)^2}-[\frac{q}{(1-f)^2}]
\end{equation}
\\ \\
Kelly, in turn, proves that,
\begin{theorem}
    The ideal fraction, which should be invested per trial under Bernoulli trials, is $g^*= p-q$, the tip. This fixed fractional approach maximizes the expected value of the capital logarithm at any court.\cite{criterion35kelly}
\end{theorem}
\begin{figure}[ht]
\centering
\includegraphics[scale=1.0]{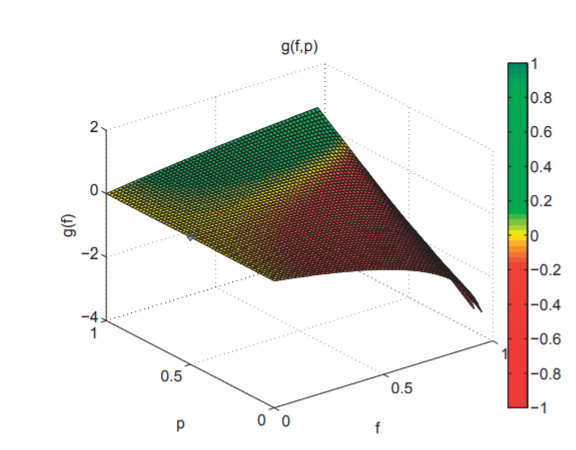}
\caption{Geometric growth rate of logarithm depending on winning and fraction probability}
\label{fig 1: Geometric growth rate of logarithm depending on winning and fraction probability}
\end{figure}
\newpage
\section{Independence and Joint Test}
In this section, hypothesis for Independence and joint test will be discussed. Basic one can be expressed using 
\\
\begin{equation}
    {H_{ q , \alpha} (\theta_0)^n_{ q= r + 1 }}
\end{equation}
\\Since the random variables of Bernoulli have serial independence so it is similar to the uncorrelation.
So, we can apply auto co-variances on equation 3.7\\
 \begin{equation}
    \Gamma_w = co-variance(H_{q,\alpha}(\theta_0), H_{q-w,\alpha}(\theta_0) \: where \: w\geq 1
\end{equation}\\
They can also be estimated by,
\\
\begin{equation}
    \Gamma_{p,w} = \frac{1}{\sqrt{p-w}} \sum^n_{q=r+w+1}(H_{q,\alpha}(\theta_0)-\alpha)(H_{q-w,\alpha}(\theta_0)-\alpha)
\end{equation}\\
Portmanteau tests can be done with the concept of Ljung \cite{ljung1978measure} and Pierce  \cite{box1970distribution} which uses the sample auto variances which later was shown by Berkowitz, Christoffersen and Pelletier. Those tests shows proper joint test with iid and unconditional hypothesis which has a marginal test of independence,  \\
\begin{equation}
   \zeta_{p,w} = \frac{1}{\sqrt{p-w}} \sum^n_{t=r+w+1}(H_{q,\alpha}(\theta_0)-E_n[H_{q,\alpha}(\theta_0)])(H_{q-k,\alpha}(\theta_0)-E_n[H_{q-w,\alpha}(\theta_0)]) 
\end{equation}
\\where, for $\theta$ $\in$ $\Theta$,
\\
\begin{align}
    E_n[H_{q,\alpha}(\theta_0)]=\frac{1}{p-w}{\sum^n_{q=r+w+1} H_{q,\alpha}(\theta_0)}
\end{align}
\\However, marginal test for (3.7) has to be on estimates of parameters which are relevant, such as,
\\
\begin{equation}
    \hat{\Gamma_{p,w}}= \frac{1}{\sqrt{p-w}} \sum^n_{q=r+w+1}(H_{q,\alpha}(\hat{\theta_0})-\alpha)(H_{q-w,\alpha}(\hat{\theta_0})-\alpha)
\end{equation}
\\or\\
\begin{equation}
 \zeta_{p,w} = \frac{1}{\sqrt{p-w}} \sum^n_{q=r+w+1}(H_{q,\alpha}(\hat{\theta_0})-E_n[H_{q,\alpha}(\hat{theta_0})]) \times (H_{q,\alpha}(\theta_0)-\hat{\alpha})(H_{q-w,\alpha}(\hat{\theta_0})-\alpha)
 \end{equation}
\\
\newpage
\section{Geometric VaR Backtesting Method}
Geometric VaR test is compressed of three individual hypothesis under a framework\cite{pelletier2016geometric}. The individual hypothesis or tests are, 
\begin{itemize}
   \item Unconditional coverage hypothesis
   \item Durable independence hypothesis
   \item Independent VaR hypothesis
\end{itemize}
it specifies the hazard function as:
\begin{equation}
    P_r{G_(t_h+n}=1 | \omega_{t_g+n-1}) = ad^{b-1}e^{cVaR_{t_g+n}}
\end{equation}
for 0 $\leq$ a $\leq$ 1 and c $\geq$ 0 , here a = unconditional coverage and $d^{b-1}$ is duration dependency. The VaR is defined on the null hypothesis and a geometrical parameter p follows, so that x = p, y=1 and z=0.
\begin{itemize}
    \item  Unconditional coverage hypothesis ( Assuming b=1 and c=0):
    \begin{align}
        H_0 : x = p \nonumber\\
        H_x : x \neq p \nonumber
    \end{align}
    
    \item Durable independence hypothesis (Assuming c=0):
    \begin{align}
        H_0 : y = 1 \nonumber\\
        H_x : y \leq 1 \nonumber
    \end{align}
    
    \item Independent VaR hypothesis:
     \begin{align}
        H_0 : z = 0 \nonumber\\
        H_x : z \geq 1 \nonumber
    \end{align}
    \item Durable independence Test and Unconditional Coverage:
    \begin{align}
        H_0 : x = p \; and \; y=1 \nonumber\\
        H_x : x \neq p  \; and \; y < 1  \nonumber
    \end{align}
    \item VaR Experiment:
    \begin{align}
        H_0 : x = p \; and \; z\:=\:0 \nonumber\\
        H_x : x \neq p  \; and \; z\: > \:0 \nonumber
    \end{align}
    \item Geometric VaR Experiment:
    \begin{align}
        H_0 : x = p, \; y=1\;  and \; z\:=\:0 \nonumber\\
        H_x : x \neq p,  \; y< 1\; and \; z\: > \:0 \nonumber
    \end{align}\\
\end{itemize}
This procedure also tests if VaR is generally misspecified.
The test statistic can be illustrated as follows:
\begin{align}
    LR = LR^{UC} + LR^{Dind} + LR^{Vind}
\end{align}\\
Where,
\begin{align}
    LR^{UC} = -2[ln\:L(x=p, y=1, z=0)] - ln\:L(\hat{x}, y=1, z=0)]\\
    LR^Dind = -2[ln \: L (\hat{x},y=1, z=0)-ln\:L(\hat{x}, \hat{y}, z=0)]\\
    LR^Vind = -2[ln\:L(\hat{x},\hat{y},\hat{z})-ln\:L(\hat{x},\hat{y}, z=0)]
\end{align}
Where L is a likelihood function. The Geometric VaR test is more powerful than any other test.

\section{Markov's Model Test}
Assume that the distribution of $I_q(\alpha)$ is conditional on a series and ultimately forms a Bernoulli Distribution from which parameters can be found.
\begin{equation}
    f_q(\theta)= n(1)_{q-1.PE_1}+...+n(m)_{q-1PE_k}+[1-\sum^m_{i=1}n(i)_{q-1}]fs
\end{equation}
\\Where,\\
\begin{equation}
    n(1)_{q-1} = I{I_{q-1}=1},....,n(m)_{q-1} = I{I_{q-1}=0,....,I_{q-m}==1}
\end{equation}
\\
\begin{equation}
  FE_1 = fr(I_q = 1 | I_{q-1}=1),...,FE_m=fr(I_q = 1 | I_{q-1}=0,...,I_{q-k}=1)
\end{equation}
\\and 
\begin{align}
    fs=Fr(I_t = 1 | I_{q-1} =0 ,...., I_{q-m}=0)\\
    under \; the \; independence,\; FE = ... = fE_k = fs= \phi \nonumber
\end{align}
\\
The paper also presents the following methods for testing conditional and independent coverage:\\
\begin{align}
    LR_1 = -2[ln(1-\hat{\sigma})(Q_{0,0} + Q_{1,0})ln\widehat{\sigma}(Q_{0,1}+Q_{1,1})-ln(1-\widehat{fs})Q_{0,0} \nonumber
    \\
    -ln\widehat{fs}Q_{0,1}-\sum^m_{i=1}ln(1-\widehat{FE_i})T_{1,0}(i) - \sum^m_{i=1}ln\widehat{FE}Q_{1,1}(i)]
\end{align}\\
and,
\begin{align}
    LR_1 = -2[ln(1-\alpha)(Q_{0,0} + Q_{1,0})ln\alpha(Q_{0,1}+Q_{1,1})-ln(1-\widehat{fs})Q_{0,0} \nonumber
    \\
    -ln\widehat{fs}Q_{0,1}-\sum^m_{i=1}ln(1-\widehat{FE_i})Q_{1,0}(i) - \sum^m_{i=1}ln\widehat{FE}Q_{1,1}(i)]
\end{align}\\ 
respectively, where $(\widehat{\sigma},\widehat{FE_1},......,\widehat{FE_m},\widehat{FS})$ are highest figure of $Q_{1,1}, Q_{0,1}, Q_{1,0}, Q_{0,0}$.\\
It was found that generalized Markov model test is better than Markov test of Christoffersen(1998) when it comes to size properties \cite{pajhede2015backtesting}.
\\ \\ 
\section{Backtesting Based on the Kalman Filter}
If constant regression coefficient does not work then the best possible alternative model will be the Kalman filter model for estimating systematical risks\cite{hamilton1994time,harvey1990forecasting,wells1994variable}.
\\
Dynamical system representation with regression for modelling $\beta$ through an auto-regressive process by a noise process $\epsilon_t$ and constant variance is the soul purpose of Kalman filter. Sharpe diagonal model can be expressed using, \\
\begin{equation}
    R_{i,j}=\alpha_i + \beta_{i,t}R_{m,t} + \epsilon_{i,t}
\end{equation}
\begin{equation}
    \beta_{i,t} = T_t\beta_{i,t} + \epsilon_{i,t}
\end{equation}\\
Harvey expressed the system with more general way \cite{harvey1990forecasting},\\
\begin{equation}
    y_t = Z_tx_t+d_t+\epsilon_t
\end{equation}
\begin{equation}
    x_t = T_tx_{t-1} + c_t + \epsilon_t
\end{equation}
\\
here equation (3.27) is known as the equation of measurement and equation (3.28) is known as equation for state.\\
$R_t$ is a matrix for constant of time coefficient matrix for the return value of stocks. Matrix $S_t$ is the matrix for state transition. $d_t$ and $c_t$ can be interpreted as vector which may be known or unknown. Finally, $\epsilon_t$ is a vector for noises and disturbances and $H_t$ as covariance matrix. It can be assumed that error term for variance is constant which is Q. H, S, d and c all are time independent.
\\ The elements of the matrices and the variances of the disturbance process can be estimated by maximizing the function, \\
\begin{equation}
    log\:L=-\frac{NS}{2}\:log 2\pi \: - \frac{1}{2}\sum^S_{s=1}\:log\:F_s - \frac{1}{2}\sum^S_{s=1}\frac{v^\prime_sv_s}{F_s}
\end{equation}
\\
where $v_{s}$ is for the time $s+1$, $v_s = y_s - \hat{y}_{s}$, and $F_s$ is its variance. These $Z_s$ and $S$ can be calculated recursively using,

\begin{align}
    x_{s|s-1} = S_{x_s} + c \\
    P_{s|s-1} = SP_sS^\prime + q\\
    v_s = y_s - R_{s}x_{s|s-1}-d\\
    F_s = Z_sP_{s|s-1}Z^\prime_s + H\\
    x_s = x_{s|s-1} +\frac{P_{s|s-1}Z^\prime_sv_s}{F_s}\\
    P_s = P_{s|s-1}-\frac{P_{s|s-1}Z^\prime_sZ_sP_{t|t-1}}{F_s}
 \end{align}
 equation (3.30) to equation (3.35) is known as Kalman Filter.

\chapter{Data and Technical Indicators}
\section{Data Collection}
For the research purposes of this paper, the dataset was collected from Yahoo Finance, covering the period from January 1, 1950, to February 28, 2020. The benchmark for the dataset was SPDR S\&P 500 which as known as SPY. It is one of the benchmarks in the United States equity market, consisting of 500 large or mid-cap companies.
Given its extensive time frame, the dataset is likely to contain numerous outliers. Outliers are one of the critical challenges for anyone who works with stock market data. Thus, treating outliers was the first challenge for this research.
\section{Data Preprocessing}
Data preprocessing is crucial, especially when working with stock data, as it can be misleading when constructing portfolios. Thus, the research did not only rely only on one benchmark, it compared with other benchmark's stock data to check if the information is valid. Some companies may have started trading after the initial data collection date of January 1, 1950. Therefore, a common solution is to exclude data for that period.

 Thus ignoring the data for that period for the is the common solution. 
To get our data normally distributed, detection of outliers were very important. Thus, I applied KS test which is also known as Kolmogorov-Smirnov Test on the dataset.
\subsection{Kolmogorov-Smirnov Test}
Kolmogorov-Smirnov test is an one dimensional and non-parametric test which is meant to compare sample distribution with standard normal distribution. To run the test one needs signals and further to filter those signals. In this research, signals were, 
\begin{table}[ht!]
\centering
\begin{tabular}{|c|c|}
\hline
signal & condition              \\ \hline
-1     & low\textgreater{}close \\
1      & low\textless{}close    \\
0      & low==close             \\ \hline
\end{tabular}
\caption{Computation of long and short signal}
\label{table:1}
\end{table}

\begin{table}[ht!]
\centering
\begin{tabular}{|c|c|c|}
\hline
Tickers & ks\_values\_5 & p\_values\_5 \\ \hline
A  & 0.04916050  & 0.84794566 \\ \hline
AAL  & 0.09811723  & 0.09824362 \\ \hline
ABC  & 0.06077138  & 0.61167223 \\ \hline
ADS  & 0.09359102  & 0.12200421 \\ \hline
CERN  & 0.04982252   & 0.83633810 \\
\hline
\end{tabular}
\caption{KS Test Output}
\label{table:2}
\end{table}
I run the signal filtering for a window size of 5, 10 and 20. After the test we get the output as such for 5 days and finally using this we picked out the good tickers.
\section{Technical Indicators}
Stock data gives us many indicators which helps us to find ways to approach portfolio for backtesting. 
For this reason visualisation and analysis of technical indicators of stock data is very important. The following plots are created using the dataset that was scraped for this research.
\subsection{Relative Strength Index}
RSI is an indicator for momentum which figure out the value of market changes to be measured in the market of an asset or stock over the conditions of the action.\cite{gumparthi2017relative}
\begin{equation}
    \mu(P_u)/(\mu(P_u)+\mu(P_d))*100
\end{equation}
\\where\\
\begin{equation}
    P_u(t)=1*(P(t)-P(t-1)) \; where \; P(t)-P(t-1)>0
\end{equation}    
\begin{equation}
    P_d(t)=-1*(P(t)-P(t-1)) \; where \; P(t)- P(t-1)<0
\end{equation}
\begin{figure}
    
    \includegraphics[width=16cm, height=20cm]{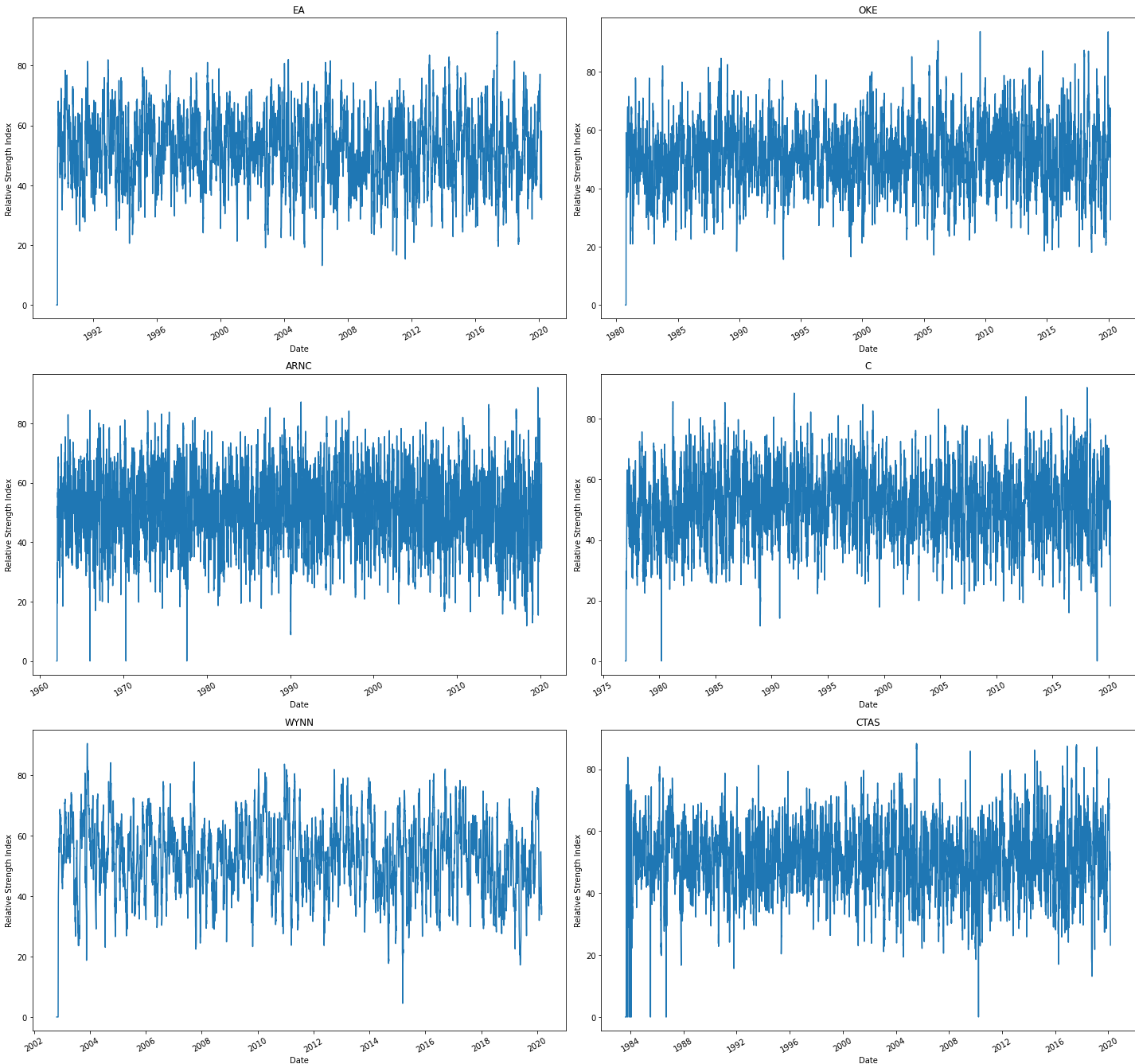}
    \caption{RSI plot}
    \label{fig 2:RSI}
\end{figure}
\newpage   
\subsection{Bollinger Bands}
Bollinger Bands was developed by John Bolliger\cite{bollinger2002bollinger}. Bollinger Bands allows to determine whether the quality is relative high or small. The method describes it using pair of lowerbound and upperbound and thus it is named as Bollinger Bands.\\  To calculate Bollinger Bands, SMA Must be calculated. Ideas of the BB was used to develop the backtest trading algorithm for this research. For lowerbound, subtraction of the standard deviation is required. For upperbound we need to add the standard deviation. 
\begin{figure}[h]
    \includegraphics[width=16cm, height=18cm]{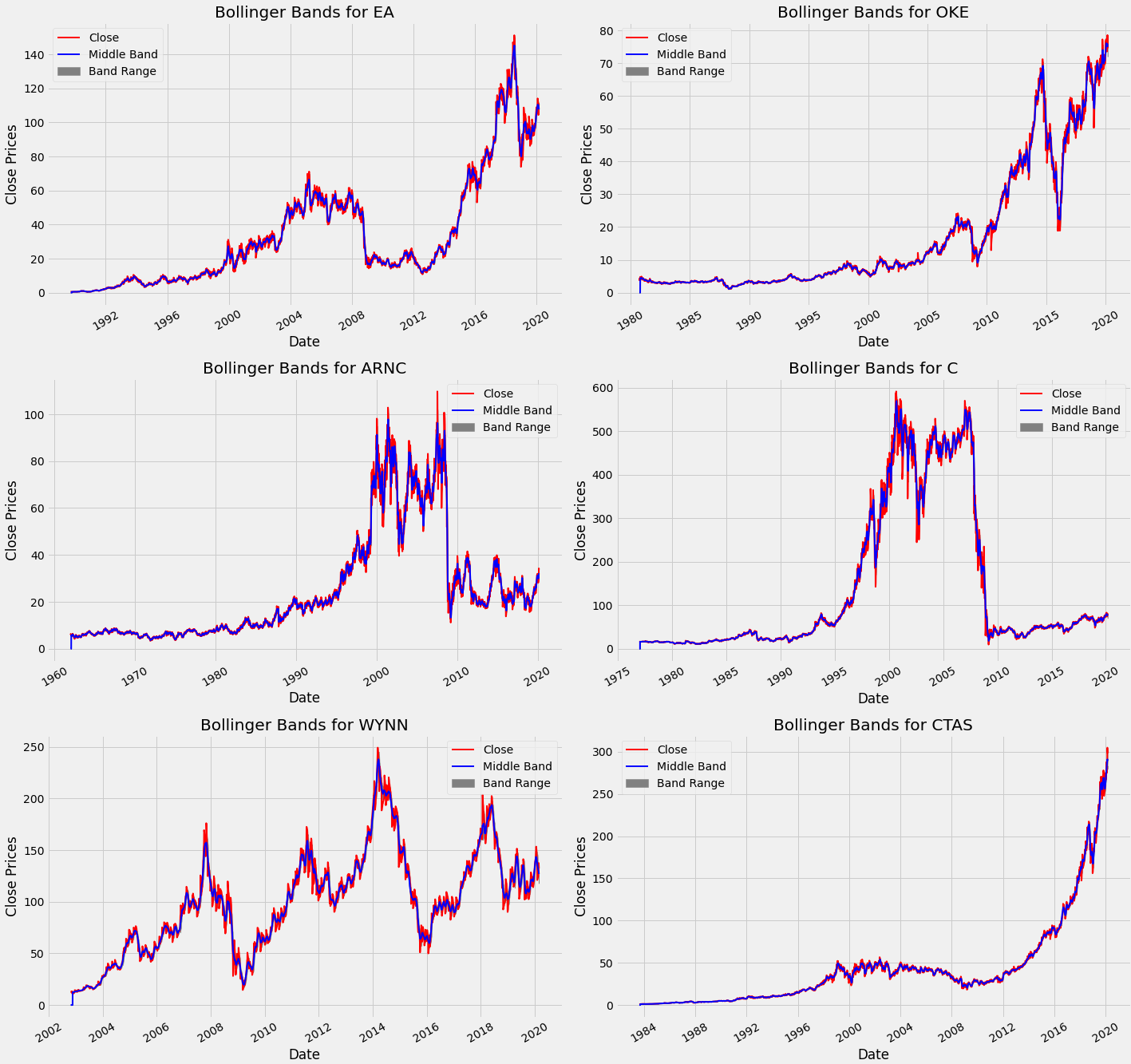}
    \caption{Bollinger Bands plot}
    \label{fig 3:Bollinger Bands}
\end{figure}
\newpage 

\subsection{SMA or Simple Moving-Average}
Throughout the timeframe of the stock data, this method calculates the average values which moves through a window size \cite{johnston1999some}. The window size can be for days, months, year or even hours or minutes. It helps to analyze the changes of price of stock data dynamically.
\begin{figure}[h]
    \centering
    \includegraphics[width=14cm, height=12cm]{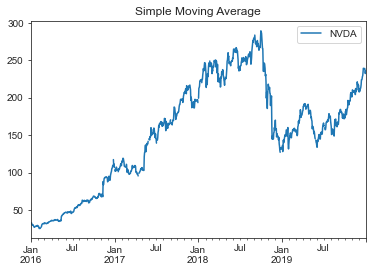}
    \caption{Simple Moving Average for NVDIA}
    \label{fig 4: Simple Moving Average for NVDIA}
\end{figure}
\newpage 
\begin{figure}[h]
    \includegraphics[width=14cm, height=12cm]{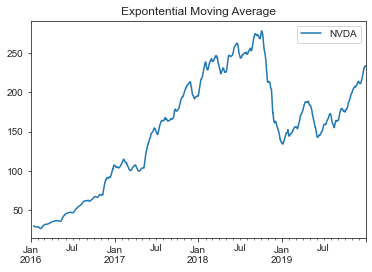}
    \caption{ Exponential Moving Average for NVDIA}
    \label{fig 5: Exponential Moving Average for NVDIA}
\end{figure}
\subsection{Aroon Oscillator}
Aroon Oscillator is a indicator for trends which uses Aroon Up and Aroon Down to measure the strength of a trend and the probaability of how it will continue \cite{chande1994new}. It uses last 25 period of high and low values to calculate Aroon up and Aroon low. The higher the indicator value the better the trend. While the oscillator moves above the x axis then we can say that the Aroon up is crossing above the Aroon down and vice versa.
\newpage
\begin{figure}[h]
    \includegraphics[width=16cm, height=15cm]{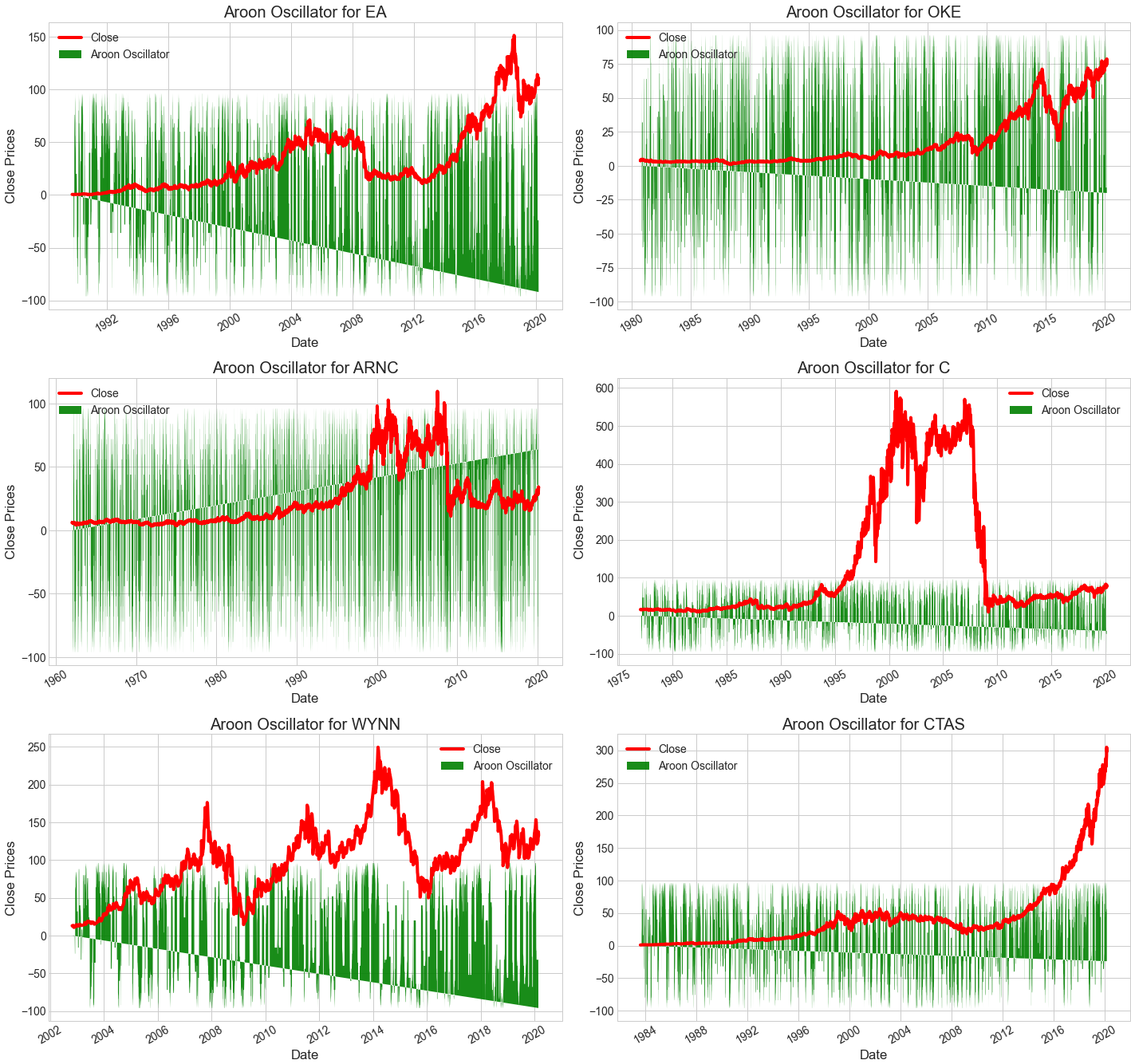}
    \caption{Aroon Oscillator plot}
    \label{fig 6: Aroon Oscillator}
\end{figure}

\subsection{Price Volume Trend (PVT)}
Price Volume Trend is also known as Volume Price Trend. It is a secure indicator which determines the securrity of price change and direction\cite{batten1996technical}. There are three types of indicators of PVT,
\begin{itemize}
    \item Single Line Crossover: just a Moving Average indicator.
    \item Confirmations: Uses conjunction with Moving Average and ADX.
    \item Divergence: It spots technical divergence.
\end{itemize}\newpage
\begin{figure}[h]
    \includegraphics[width=16cm, height=18cm]{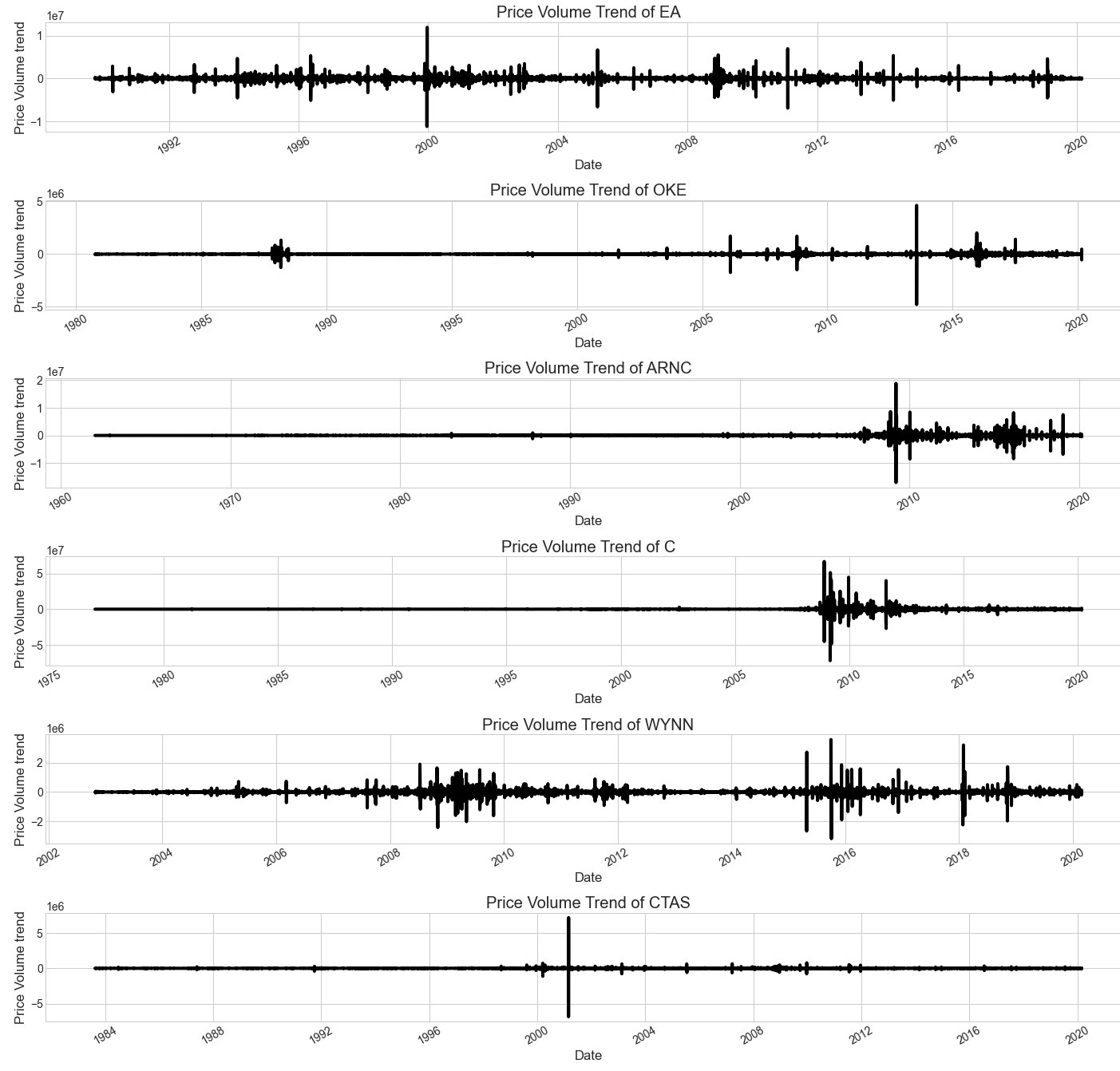}
    \caption{Price Volume Trend}
    \label{fig 7: Price Volume Trend}
\end{figure}
\subsection{Acceleration Bands}
The idea is to get into a trade when the security trends highly but before it's price moves drastically in any direction. It measures volatility on number of bars which is defined by the traders\cite{headley2002big}. If price goes upper than top band then it is called Accelerated. It can be interpreted as eit for long position. 
\newpage
\begin{figure}[h]
    \includegraphics[width=16cm, height=18cm]{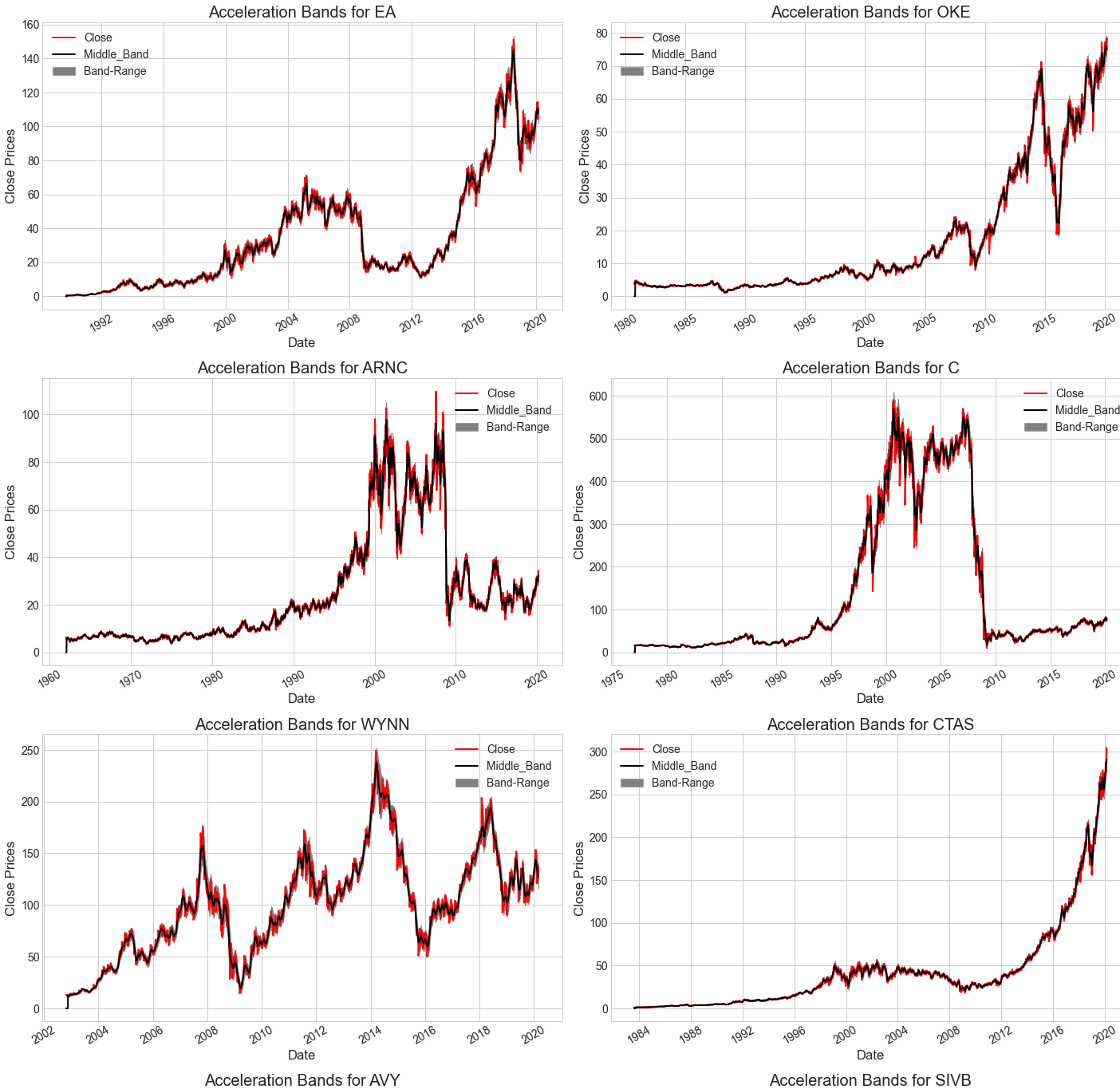}
    \caption{Accelaration Bands plot}
    \label{fig 8: Accelaration Bands plot }
\end{figure}
\subsection{Stochastic Oscillator}
Stochastic Oscillator is a measurement of momentum that compares the closing price of a security of a given stock to the price of its stock over a timeframe \cite{burgers1999nino}.
The formula can be expressed as,
\begin{equation}
    \%k = (\frac{c-l_{4}}{h_{4}-l_(4)})\times 100
\end{equation}
Where, \\
c = Recent closing price\\
$l_{4}$ = Lowest price in last 4 days\\
$h_4$ = Highest price in last 4 days\\
\%k = The current value of stochastic oscillator\\

\begin{figure}[h]
    \includegraphics[width=16cm, height=16cm]{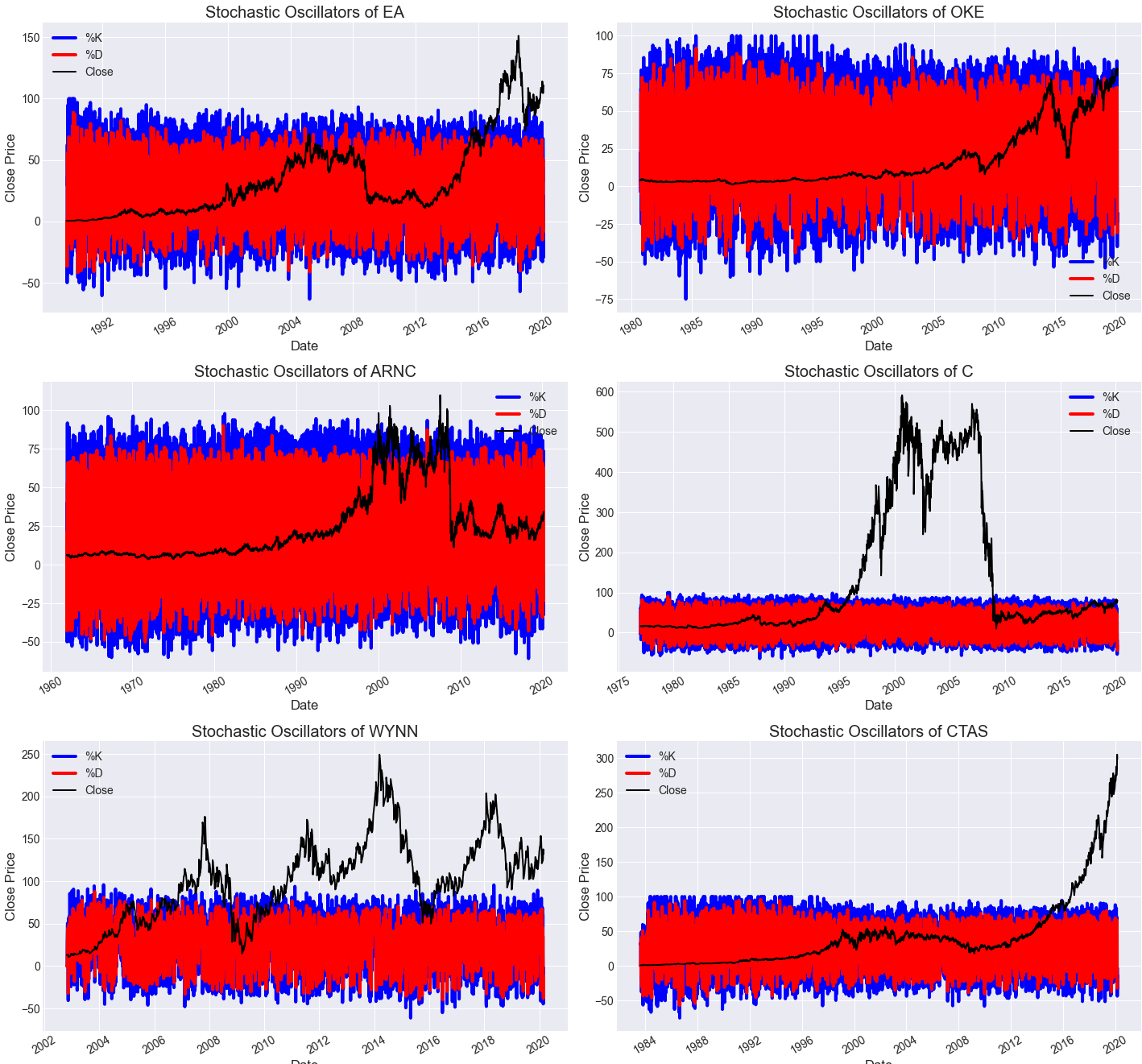}
    \caption{Stochastic Oscillator plot}
    \label{fig 9: Stochastic Oscillator plot}
\end{figure}

\subsection{Chaikin Money Flow}
Chaikin Money Flow or CMF is an average of volume-weighted maginitude of distribution and accumulation for a specific time period\cite{thomsett2010cmf}. It was developed by Marc Chaikin. 
\begin{itemize}
    \item If the CMF value is above x axis then it can be considered as the strength of the market and vice versa.
    \item When CMF diverging with a value of higher low, it begins to rise with lower low oversold zone.
    
\end{itemize}
\begin{figure}[h]
    \includegraphics[width=16cm, height=18cm]{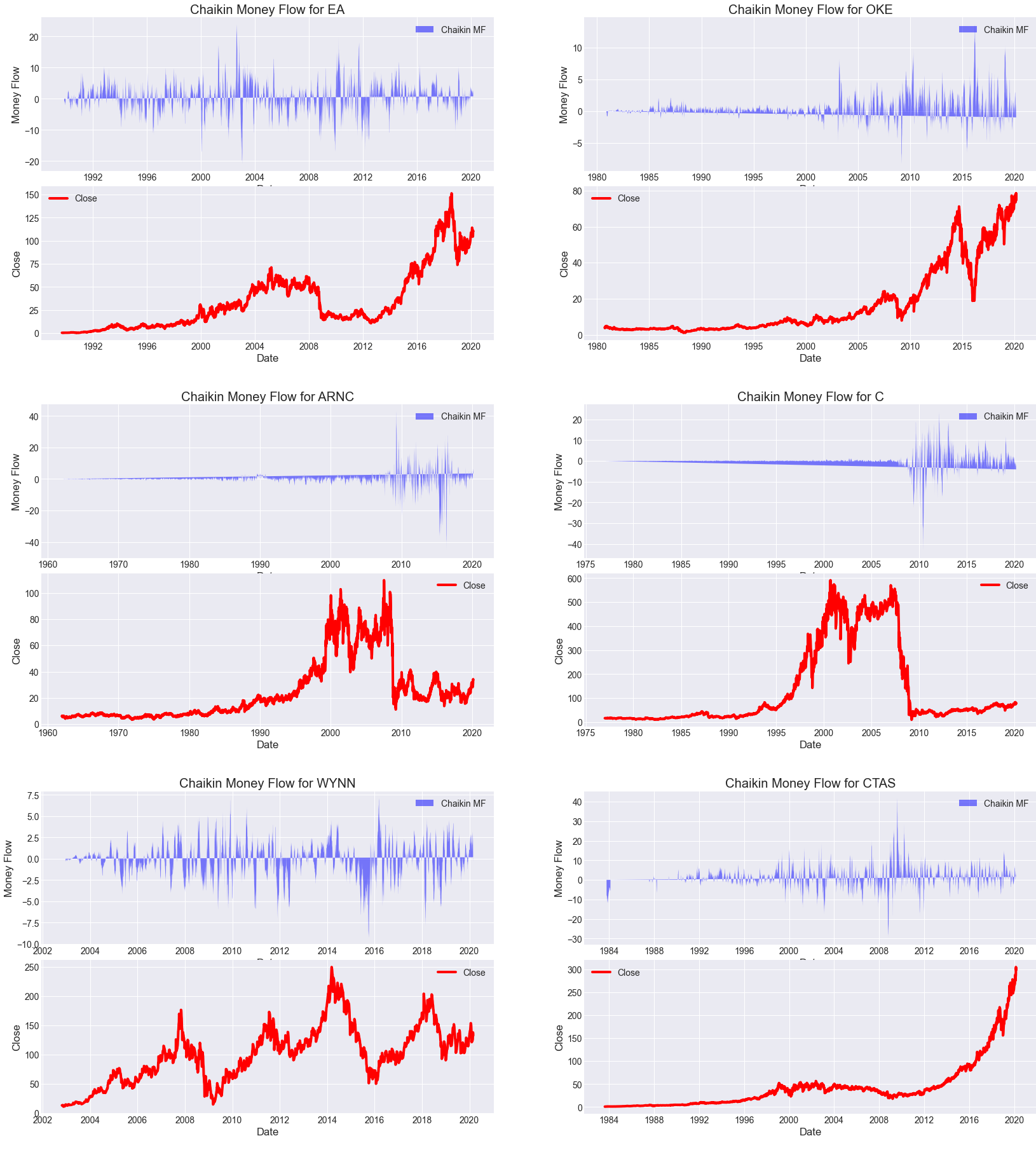}
    \caption{Chaikin Money Flow plot}
    \label{fig 10: Chaikin Money Flow plot}
\end{figure}
\newpage

\subsection{Parabolic SAR} 
This indicator is quite interesting as it provides entry and exit size along with the high-lightened direction for the moving asset\cite{wilder1978momentum}. The indicator suggests price direction for a stock. It is also known as "Stop and Reversal System".
\begin{figure}[h]
    \includegraphics[width=16cm, height=18cm]{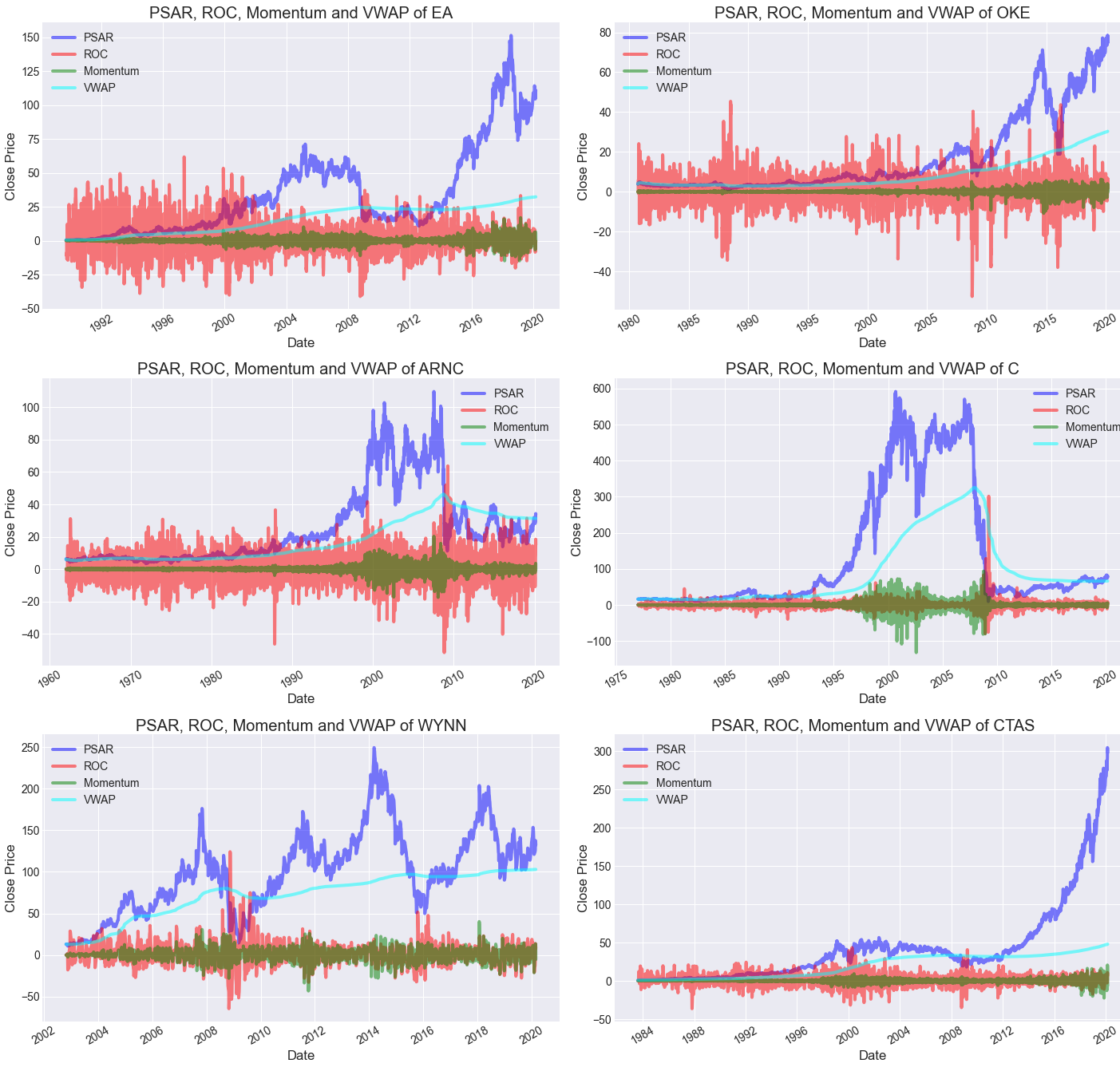}
    \caption{Parabolic SAR plot}
    \label{fig 11: Parabolic SAR plot}
\end{figure}
\newpage
\subsection{Klenter Channels}
It is more like bands indicators which set above and below of an exponential moving average to indicate volatility\cite{evens1999keltner}. It uses ATR to set distance for channels. 
\begin{figure}[h]
    \includegraphics[width=16cm, height=18cm]{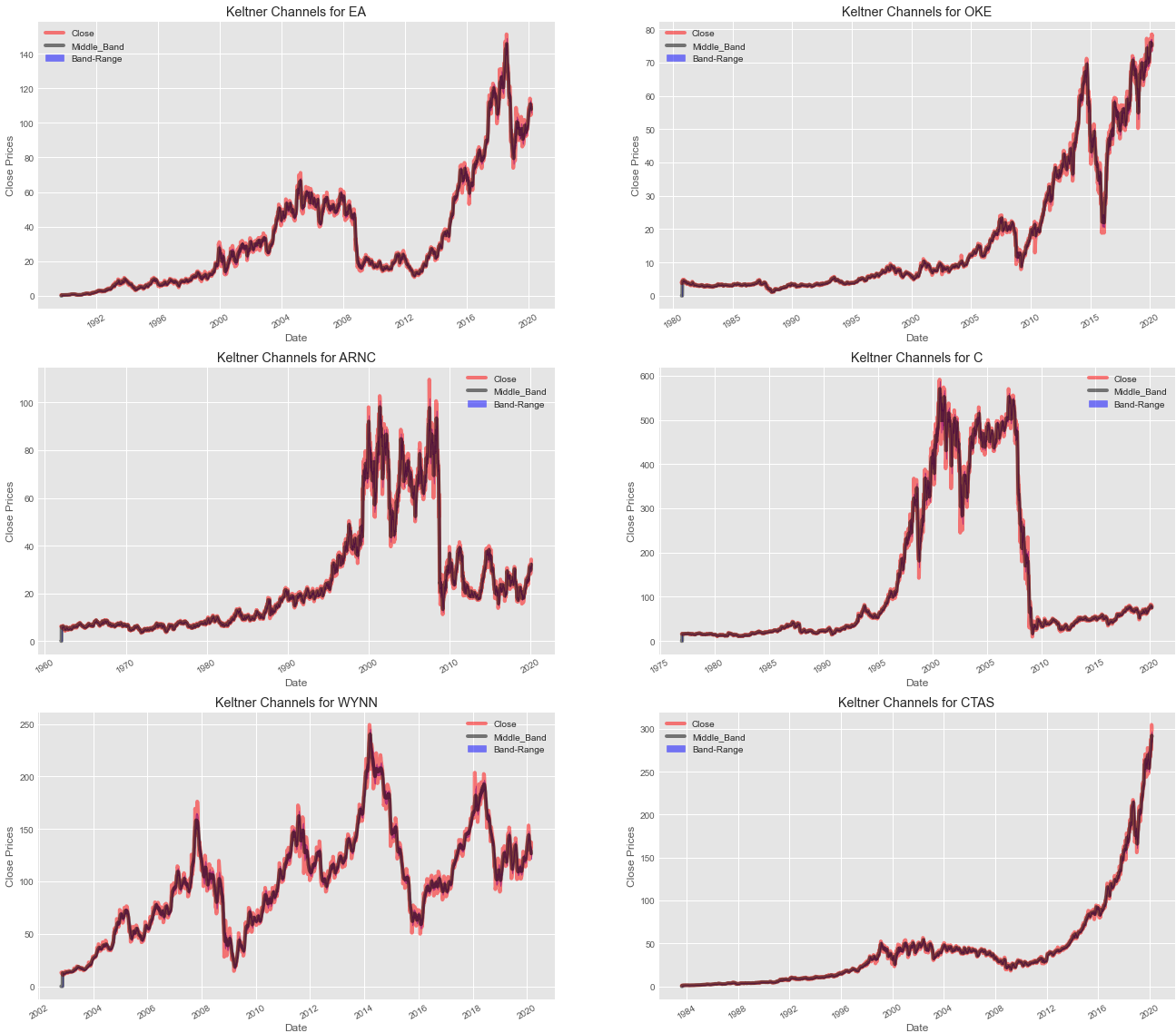}
    \caption{Kelnter Channel plot}
    \label{fig 12: Kelnter Channel plot}
\end{figure}
\newpage
\subsection{Ichimoku kinko Cloud}
This is a critical signal that distinguishes resistance and support and identifies trend path, provides signal for trading and gauges momentum \cite{elliott2007ichimoku}. It is know n as "One Look Equilibrium Chart"
\begin{figure}[h]
    \includegraphics[width=16cm, height=18cm]{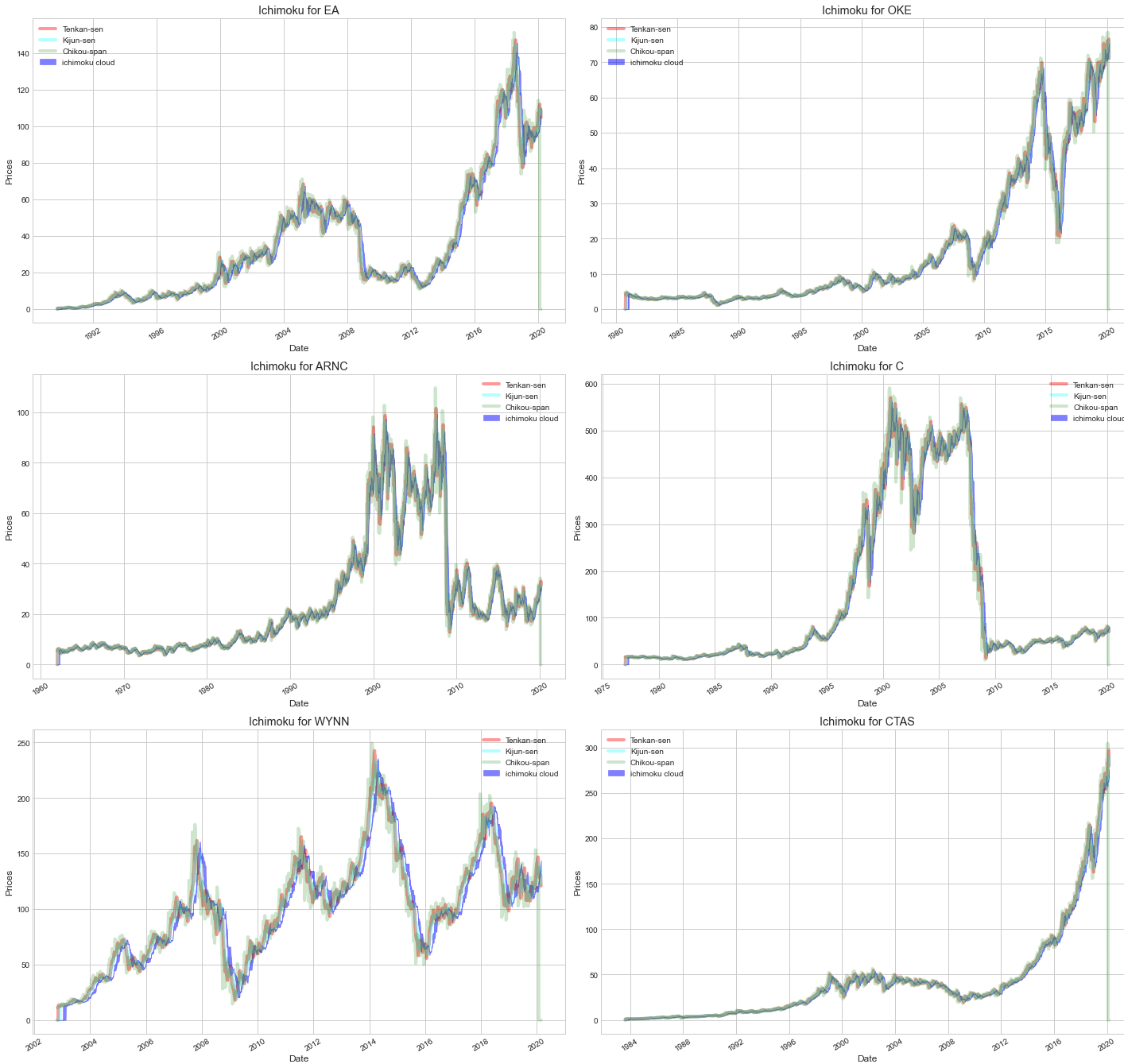}
    \caption{Ichimoku Kinko Cloud plot}
    \label{fig 13: Ichimoku Kinko Cloud plot}
\end{figure}\newpage
\subsection{Money Flow Index}
It is actually a oscillator that identifies over sold or overbought conditions in asset while spotting divergence in trend \cite{tharavanij2017profitability}. The oscillator moves in between 0 to 100. The equations can be expressed as,
\begin{align}
    Price = (High + Low + Close)/3.0
    \\Raw Money Flow = Price x Volume\\
    Money Flow Ratio = (14 period +ve Money Flow)/(14 period -ve Money Flow)
    \\Money Flow Index = 100 - \frac{100}{(1 + Money Flow Ratio)}
\end{align}\newpage \clearpage
\begin{figure}[h]
    \includegraphics[width=16cm, height=23cm]{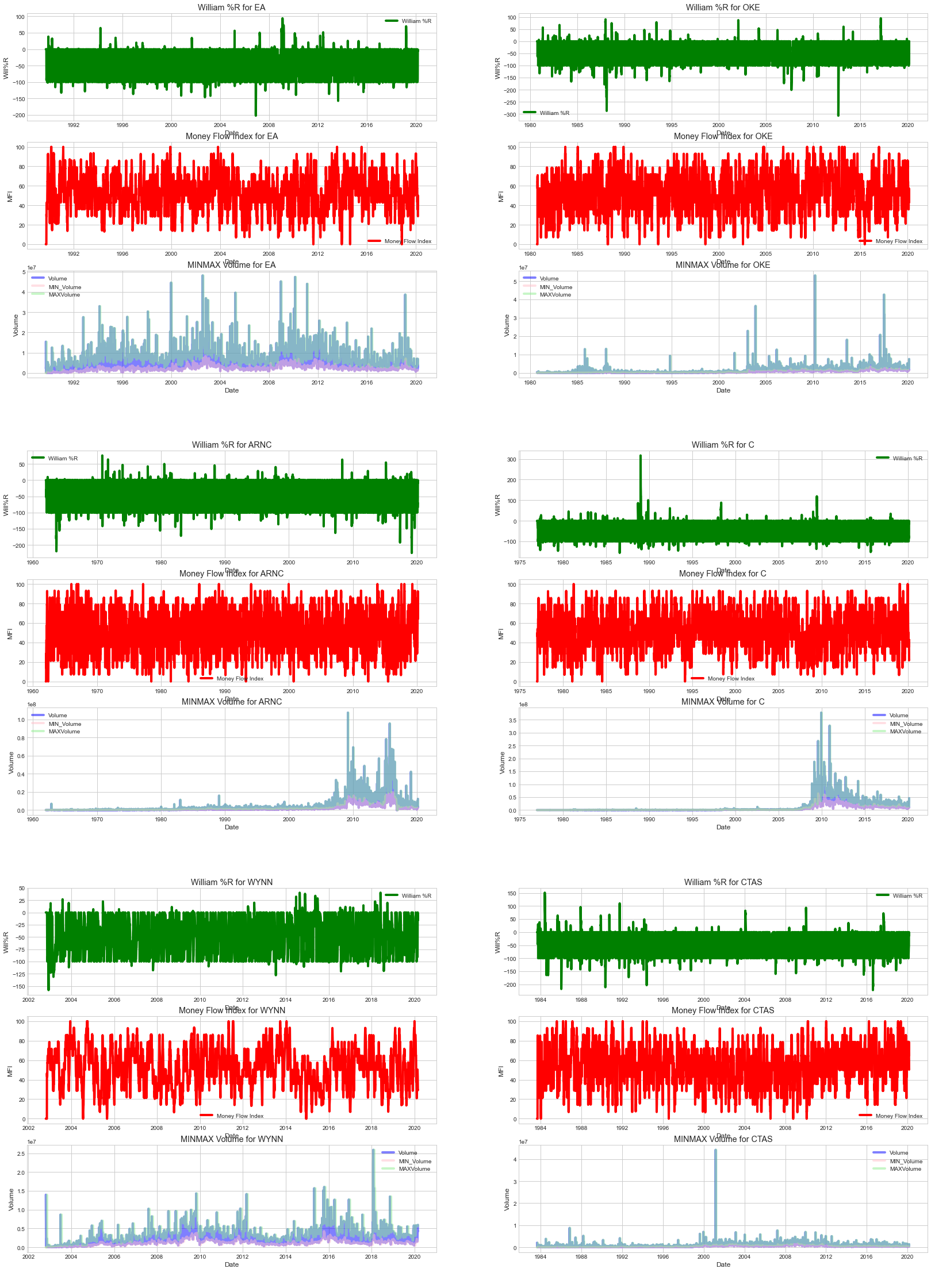}
    \caption{Money Flow Index plot}
    \label{fig 14: Money Flow Index plot}
\end{figure}
\newpage \clearpage
\subsection{Correlation Between Columns}
\begin{figure}[!htb]
   \begin{minipage}{0.6\textwidth}
     \centering
     \includegraphics[width=.6\linewidth]{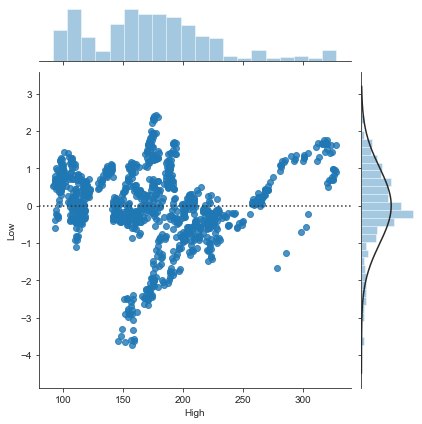}
     \caption{High price and low price link}\label{Fig 15:Correlation between high and low price}
   \end{minipage}\hfill
   \begin{minipage}{0.6\textwidth}
     \centering
     \includegraphics[width=.6\linewidth]{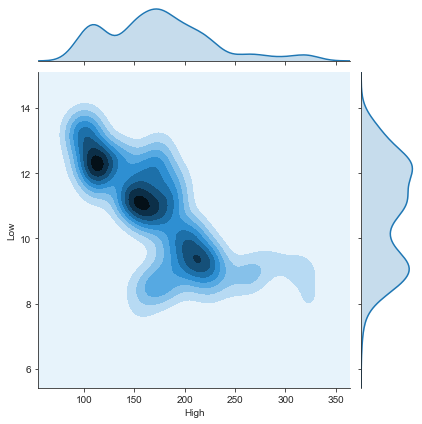}
     \caption{High price and low price link (modified)}\label{Fig 16: Correlation between high and low price mod}
   \end{minipage}
\end{figure}
\begin{figure}[!htb]
   \begin{minipage}{0.6\textwidth}
     \centering
     \includegraphics[width=.6\linewidth]{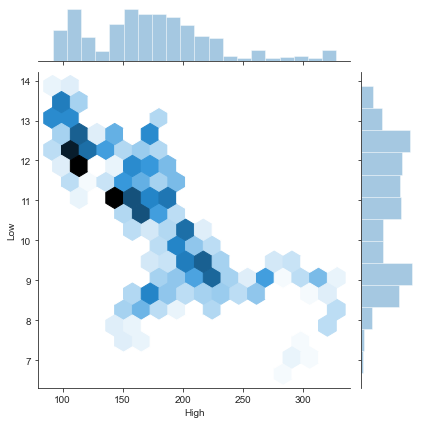}
     \caption{High price and low price link (second modification)}\label{Fig 17:Correlation between high and low price second modification }
   \end{minipage}\hfill
   \begin{minipage}{0.6\textwidth}
     \centering
     \includegraphics[width=.6\linewidth]{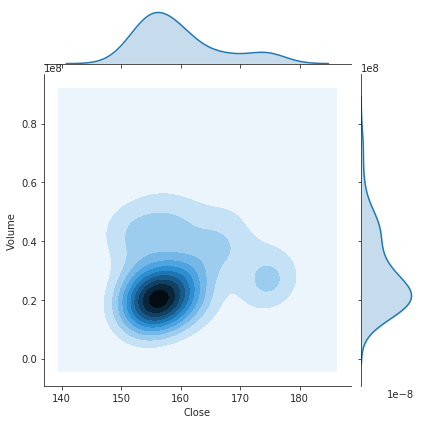}
     \caption{volume and close price link}\label{Fig 18:Correlation between volume and close price}
   \end{minipage}
\end{figure}

\chapter{Result of Backtesting Analysis}
\section{Alpha Factor Analysis}
Building upon the technical indicators analyzed in the previous chapter, I developed my own trading algorithm for backtesting, incorporating similar factors. The alpha factor can be defined as the excess returns gained compared to benchmark returns. Beta, on the other hand, represents a factor that exhibits multiplicative characteristics. Beta actually measures the relative volatility. For this analysis, 10 quantiles over three different time periods: 1, 5, and 10 days were defined. Below is an analysis of the factors utilized in the backtest.

\begin{figure}[h]
    \centering
    \includegraphics[scale=1]{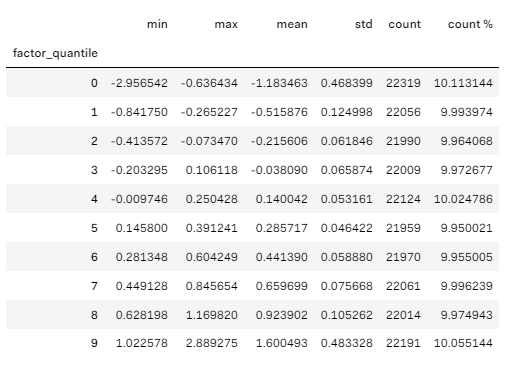}
    \caption{Quantile analysis of the factors at 20 percent}
    \label{fig 19: Quantile analysis of the factors at 20 percent}
\end{figure}

\clearpage

\begin{figure}[h]
    \centering
    \includegraphics[scale=1]{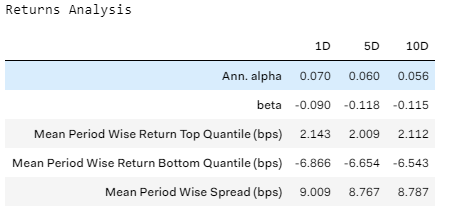}
    \caption{Study of factor return }
    \label{fig 20: Study of factor return}
\end{figure}

\begin{figure}[h]
    \centering
    \includegraphics[scale=1]{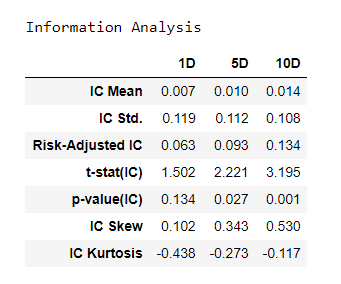}
    \caption{Study of the information of the factors }
    \label{fig 21: Information analysis of the factors}
\end{figure}

Figure 5.4 below illustrates that our mean factor values are favorable, and the deviation, often interpreted as volatility, is also well within acceptable bounds. So, the volatility ensures that our factors will perform well.

\begin{figure}[h]
    \centering
    \includegraphics[width=17cm, height=18cm]{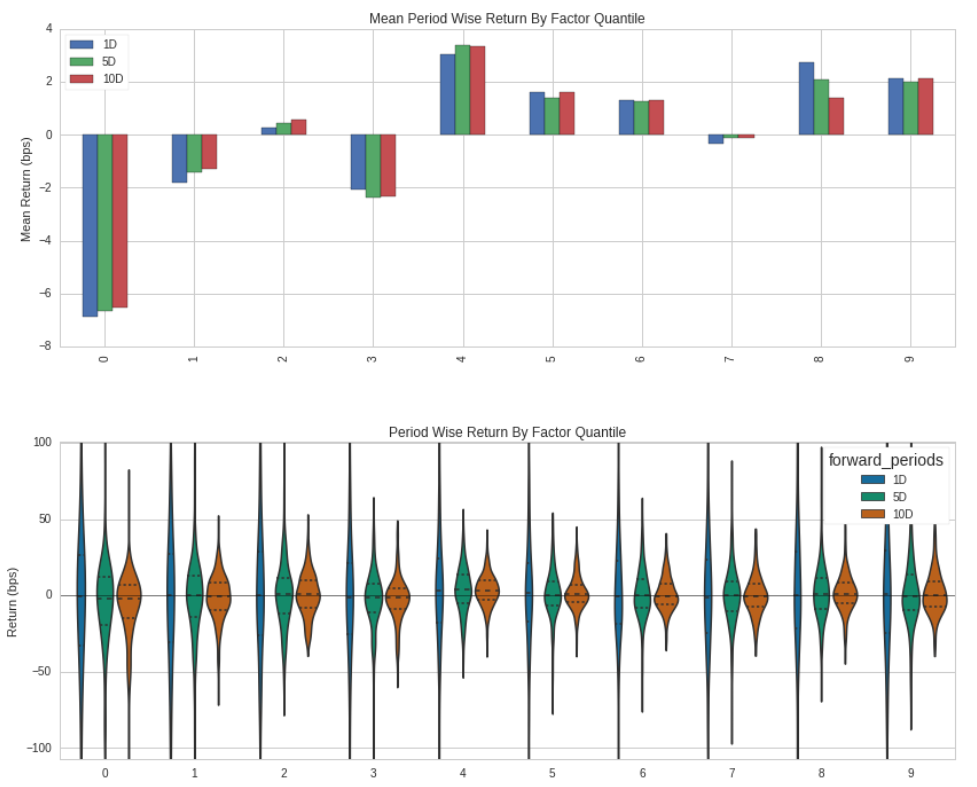}
    \caption{Mean and Volatility analysis of the factors }
    \label{fig 22: Mean and Volatility analysis of the factors}
\end{figure}

\clearpage

\begin{figure}[h]
    \centering
    \includegraphics[width=17cm, height=8cm]{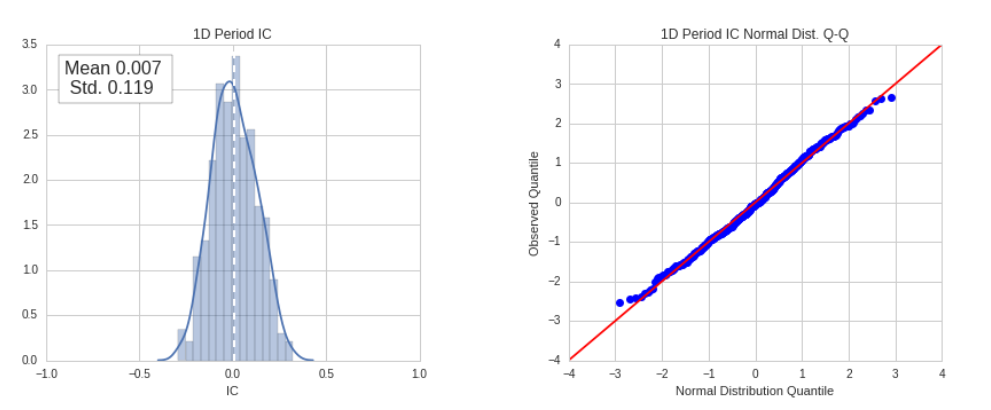}
    \caption{1 day period IC and Normal Q-Q distribution}
    \label{fig 23: Information analysis of the factors}
\end{figure}

\begin{figure}[h]
    \centering
    \includegraphics[width=17cm, height=8cm]{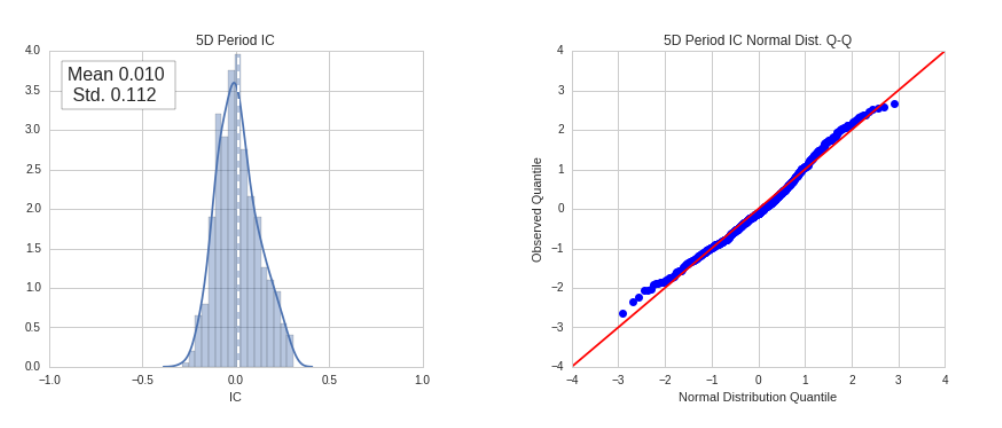}
    \caption{5 day period IC and Normal Q-Q distribution}
    \label{fig 24: Information analysis of the factors}
\end{figure}

\clearpage

\begin{figure}[h]
    \centering
    \includegraphics[width=17cm, height=8cm]{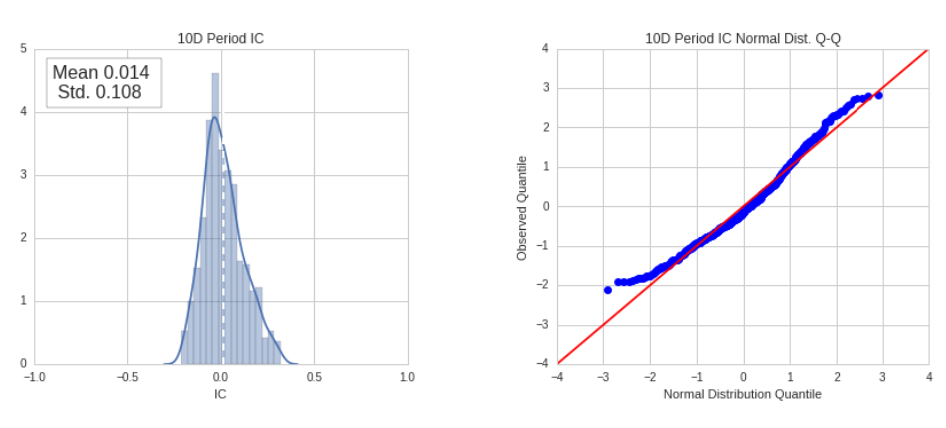}
    \caption{5 day period IC and Normal Q-Q distribution}
    \label{fig 25: Information analysis of the factors}
\end{figure}

\section{Backtesting Result}
Here, some definitions of result analysis metrics are provided:

\begin{itemize}
    \item Total Returns: The total  percentage return of the portfolio from the start to the end of the backtest.
    \item Specific Returns: The difference between the portfolio's total returns and common returns.
    \item Sharpe Ratio: A risk-adjusted performance metric, which divides the return of the portfolio which is excess over the risk-free rate by the standard deviation of the portfolio.
    \item Sortino Ratio: As same as Sharpe ratio but it does not count with short positional outcomes.
    \item Max-Draw Down: The largest peak-to-trough drop in the portfolio's history.
    \item Volatility: Standard deviation or $\sigma$ of the returns of the portfolio.
\end{itemize}

\clearpage

\subsection{Independence and Joint Test}
In this section, the paper will illustrate the independence and joint test using different long and short positional values.

\subsubsection{Hedging with positional value at 10 percent}

\begin{figure}[h]
    \centering
    \includegraphics[width=4cm, height=3cm]{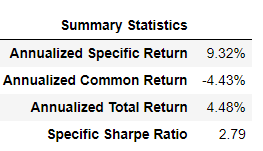}
    \caption{Performance Relative to Risk Factor}
    \label{fig 26: Performance Relative to Risk Factor}
\end{figure}

\begin{figure}[!htb]
    \centering
    \includegraphics[width=8cm, height=9cm]{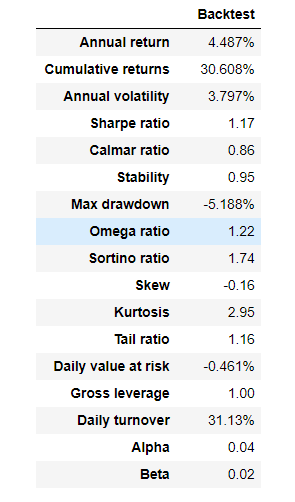}
    \caption{Backtest Information with hedging}
    \label{fig 27: Backtest Information with hedging}
\end{figure}

\begin{figure}[!htb]
    \centering
    \includegraphics[width=17cm, height=5cm]{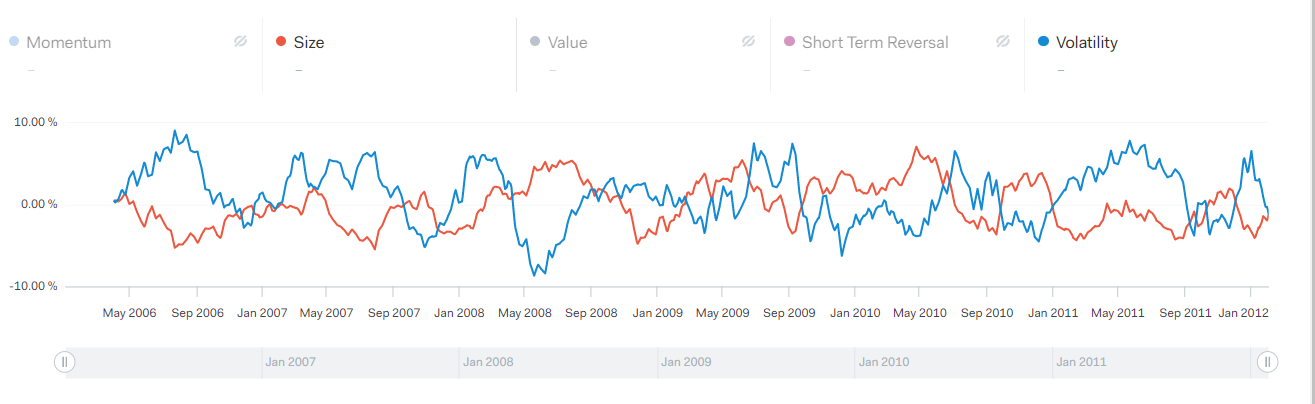}
    \caption{Size and Volatility Comparison for Hedge}
    \label{fig 28: Size and volatility comparison for hedge}
\end{figure}

\begin{figure}[!htb]
    \centering
    \includegraphics[width=17cm, height=5cm]{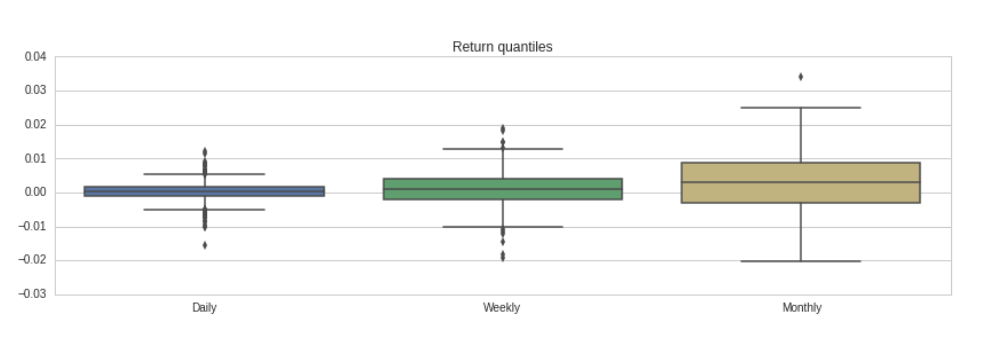}
    \caption{Return Quantiles over Day, Month, Week for Hedge}
    \label{fig 29: Return Quantiles over Day, Month, Week for Hedge}
\end{figure}

\clearpage

\begin{figure}[!htb]
    \centering
    \includegraphics[width=17cm, height=5cm]{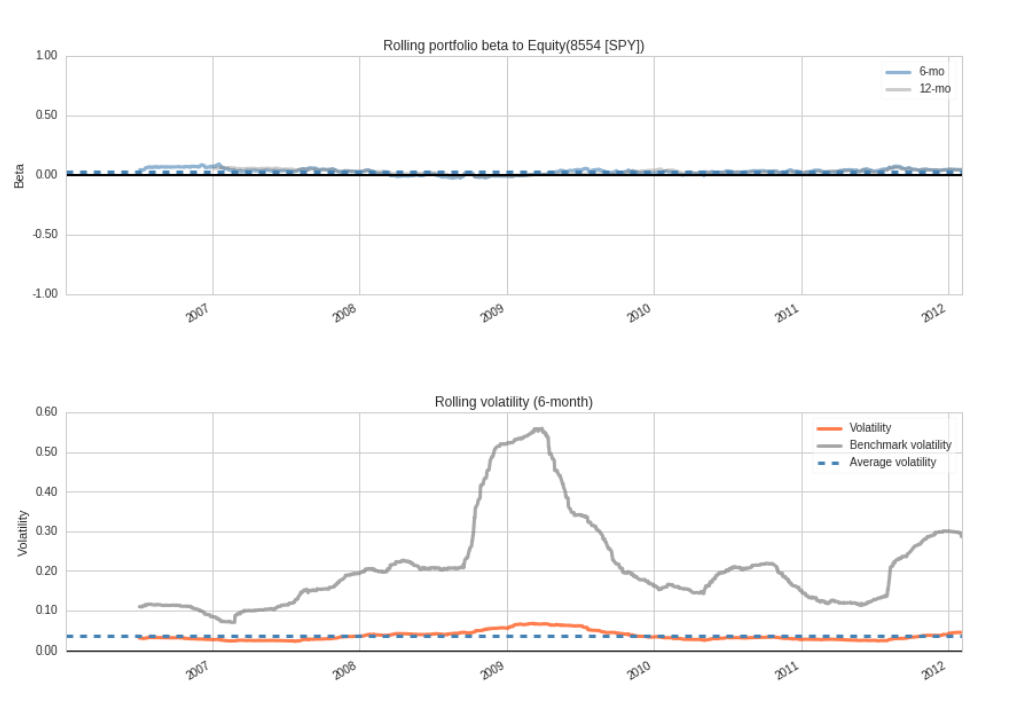}
    \caption{Rolling Volatility for Hedges}
    \label{fig 30: Rolling Volatility for Hedges}
\end{figure}

\begin{figure}[!htb]
    \centering
    \includegraphics[width=17cm, height=5cm]{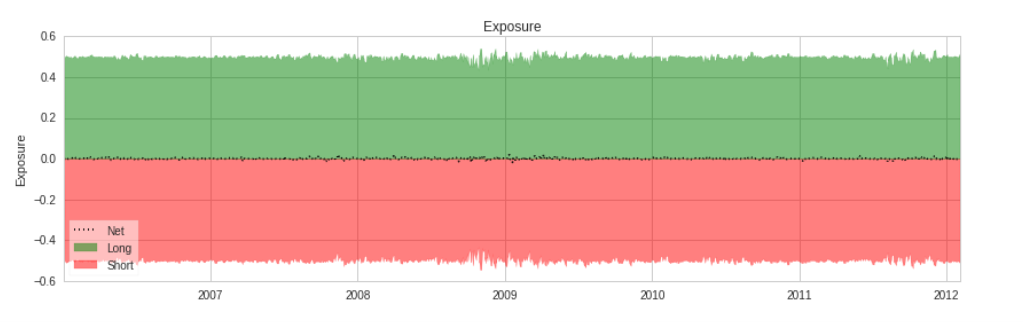}
    \caption{Exposure for long and short sizes for Hedges}
    \label{fig 31: Exposure for long and short sizes for Hedges}
\end{figure}

\begin{figure}[!htb]
    \centering
    \includegraphics[width=12cm, height=8cm]{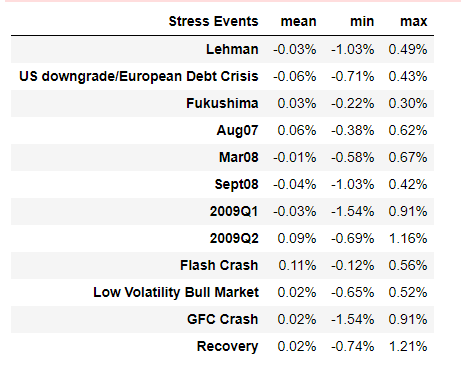}
    \caption{Performance of the backtest during high volatile situations for Hedges}
    \label{fig 32: Performance of the backtest during high volatile situations for Hedges}
\end{figure}

The maximum Value at Risk (VaR) is 0.034561 for a position size of 90.000000.

\clearpage

\subsubsection{Hedging with short positional value at 10 percent and long at 20 percent}

\begin{figure}[h]
    \centering
    \includegraphics[width=4cm, height=3cm]{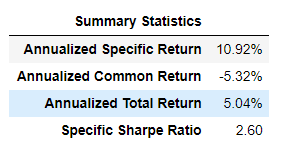}
    \caption{Performance Relative to Risk Factor short positional value at 10 percent and long at 20 percent}
    \label{fig 33: Performance Relative to Risk Factor short positional value at 10 percent and long at 20 percent}
\end{figure}

\begin{figure}[!htb]
    \centering
    \includegraphics[width=8cm, height=9cm]{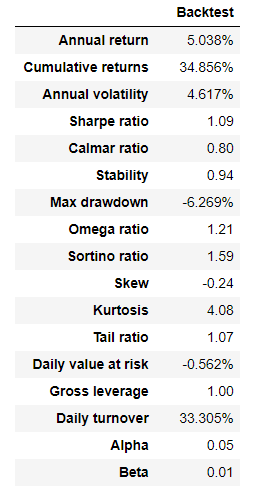}
    \caption{Backtest Information with short positional value at 10 percent and long at 20 percent}
    \label{fig 34: Backtest Information with short positional value at 10 percent and long at 20 percent}
\end{figure}

\clearpage

\begin{figure}[!htb]
    \centering
    \includegraphics[width=17cm, height=5cm]{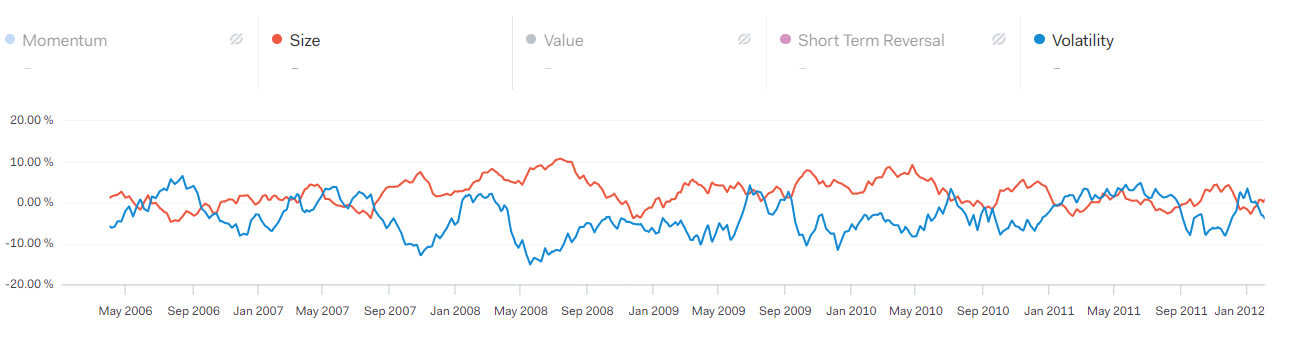}
    \caption{Size and Volatility Comparison for short positional value at 10 percent and long at 20 percent}
    \label{fig 35: Size and volatility comparison for short positional value at 10 percent and long at 20 percent}
\end{figure}

\begin{figure}[!htb]
    \centering
    \includegraphics[width=17cm, height=5cm]{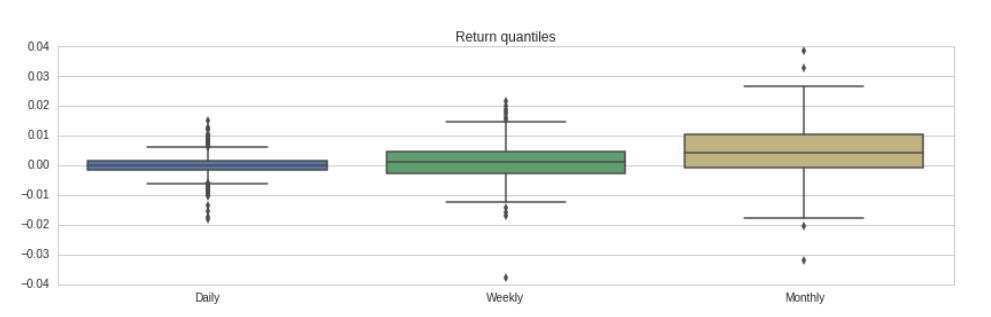}
    \caption{Return Quantiles over Day, Month, Week for short positional value at 10 percent and long at 20 percent}
    \label{fig 36: Return Quantiles over Day, Month, Week for short positional value at 10 percent and long at 20 percent}
\end{figure}

\begin{figure}[!htb]
    \centering
    \includegraphics[width=17cm, height=5cm]{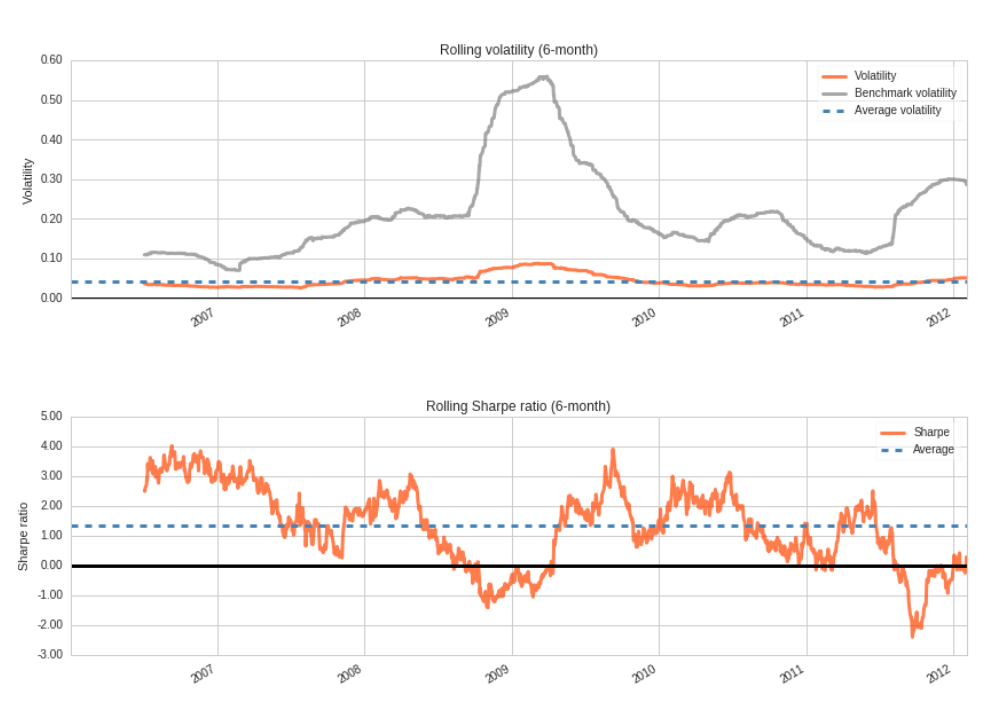}
    \caption{Rolling Volatility for short positional value at 10 percent and long at 20 percent}
    \label{fig 37: Rolling Volatility for short positional value at 10 percent and long at 20 percent}
\end{figure}

\clearpage

\begin{figure}[!htb]
    \centering
    \includegraphics[width=17cm, height=5cm]{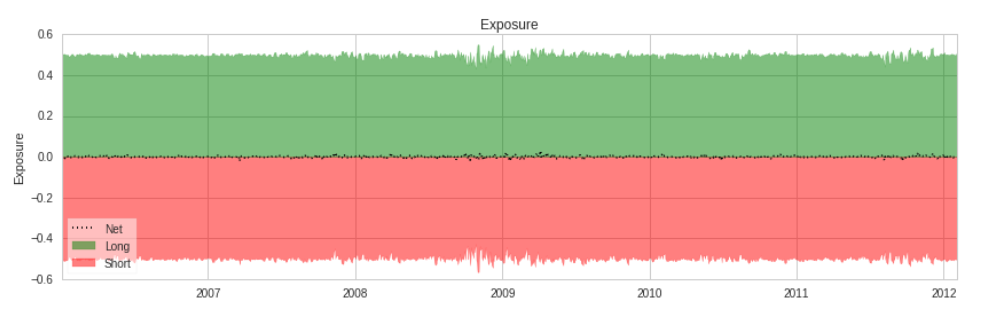}
    \caption{Exposure for long and short sizes for short positional value at 10 percent and long at 20 percent}
    \label{fig 38: Exposure for long and short sizes for short positional value at 10 percent and long at 20 percent}
\end{figure}

\begin{figure}[!htb]
    \centering
    \includegraphics[width=12cm, height=8cm]{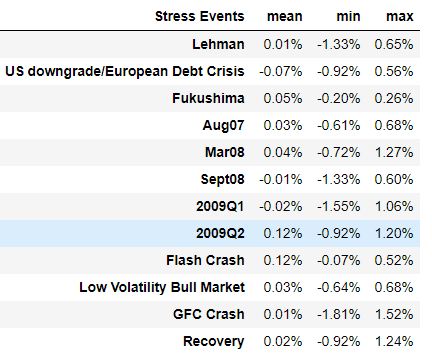}
    \caption{Performance of the backtest during high volatile situations for short positional value at 10 percent and long at 20 percent}
    \label{fig 39: Performance of the backtest during high volatile situations for short positional value at 10 percent and long at 20 percent}
\end{figure}

The maximum Value at Risk (VaR) is 0.029217 for a position size of 85.000000.

\clearpage

\subsubsection{Hedging with short positional value at 20 percent and long at 10 percent}

\begin{figure}[h]
    \centering
    \includegraphics[width=4cm, height=3cm]{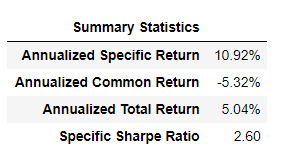}
    \caption{Performance Relative to Risk Factor short positional value at 20 percent and long at 10 percent}
    \label{fig 40: Performance Relative to Risk Factor short positional value at 20 percent and long at 10 percent}
\end{figure}

\begin{figure}[!htb]
    \centering
    \includegraphics[width=8cm, height=9cm]{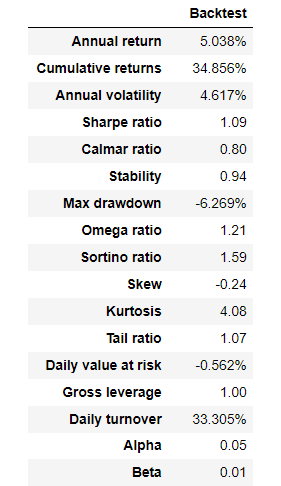}
    \caption{Backtest Information with short positional value at 20 percent and long at 10 percent}
    \label{fig 41: Backtest Information with short positional value at 20 percent and long at 10 percent}
\end{figure}

\clearpage

\begin{figure}[!htb]
    \centering
    \includegraphics[width=17cm, height=5cm]{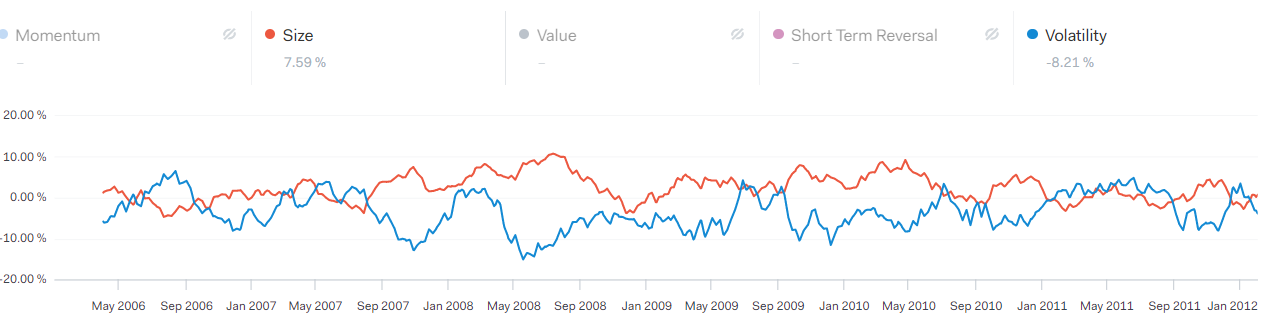}
    \caption{Size and Volatility Comparison for short positional value at 20 percent and long at 10 percent}
    \label{fig 42: Size and volatility comparison for short positional value at 20 percent and long at 10 percent}
\end{figure}

\begin{figure}[!htb]
    \centering
    \includegraphics[width=17cm, height=5cm]{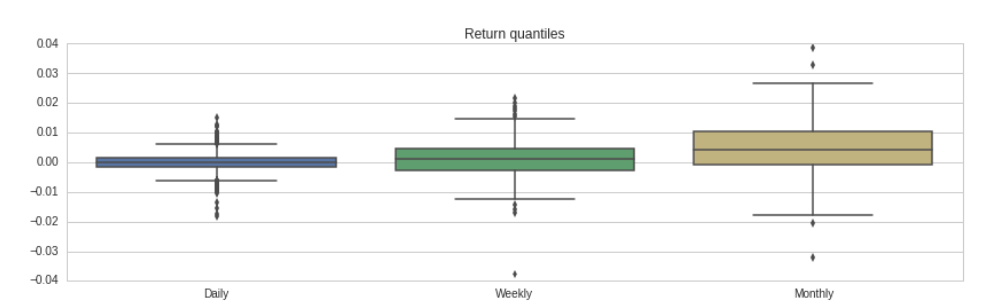}
    \caption{Return Quantiles over Day, Month, Week for short positional value at 20 percent and long at 10 percent}
    \label{fig 43: Return Quantiles over Day, Month, Week for short positional value at 20 percent and long at 10 percent}
\end{figure}

\clearpage

\begin{figure}[!htb]
    \centering
    \includegraphics[width=17cm, height=5cm]{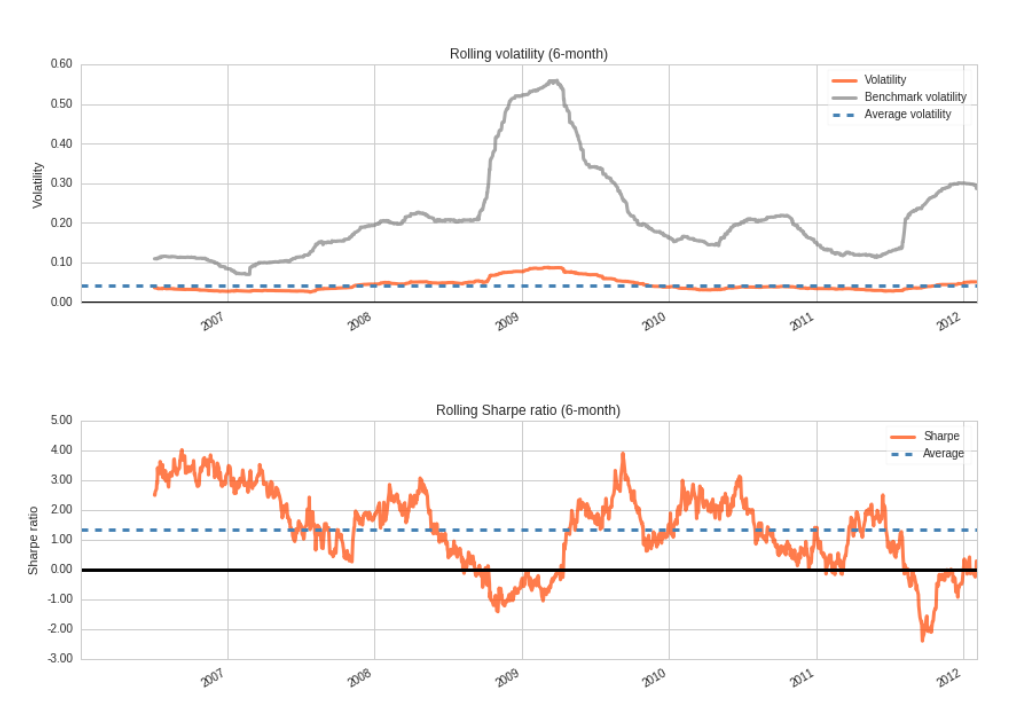}
    \caption{Rolling Volatility for short positional value at 20 percent and long at 10 percent}
    \label{fig 44: Rolling Volatility for short positional value at 20 percent and long at 10 percent}
\end{figure}

\begin{figure}[!htb]
    \centering
    \includegraphics[width=17cm, height=5cm]{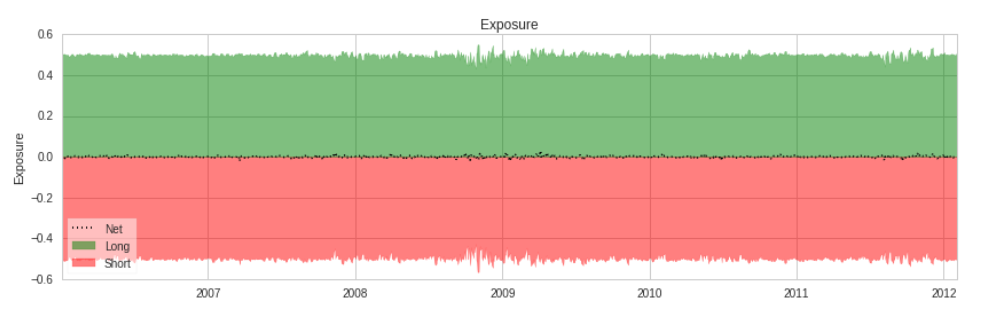}
    \caption{Exposure for long and short sizes for short positional value at 20 percent and long at 10 percent}
    \label{fig 45: Exposure for long and short sizes for short positional value at 20 percent and long at 10 percent}
\end{figure}

\begin{figure}[!htb]
    \centering
    \includegraphics[width=12cm, height=8cm]{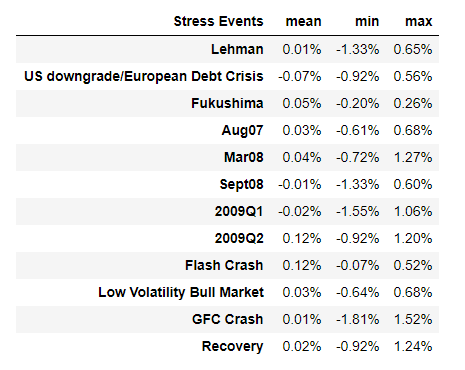}
    \caption{Performance of the backtest during high volatile situations for short positional value at 20 percent and long at 10 percent}
    \label{fig 46: Performance of the backtest during high volatile situations for short positional value at 20 percent and long at 10 percent}
\end{figure}

The maximum Value at Risk (VaR) is 0.023184 for a position size of 80.000000.

\clearpage

\subsection{Backtest using Kalman Filter Methodology}

\begin{figure}[h]
    \centering
    \includegraphics[width=4cm, height=3cm]{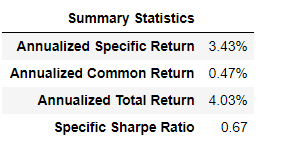}
    \caption{Performance Relative to Risk Factor Kalman Filter Methodology}
    \label{fig 47: Performance Relative to Risk Factor Kalman Filter Methodology}
\end{figure}

\begin{figure}[!htb]
    \centering
    \includegraphics[width=8cm, height=9cm]{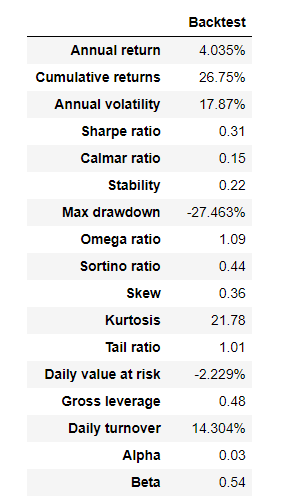}
    \caption{Backtest Information for Kalman Filter Methodology}
    \label{fig 48: Backtest Information for Kalman Filter Methodology}
\end{figure}

\clearpage

\begin{figure}[!htb]
    \centering
    \includegraphics[width=17cm, height=5cm]{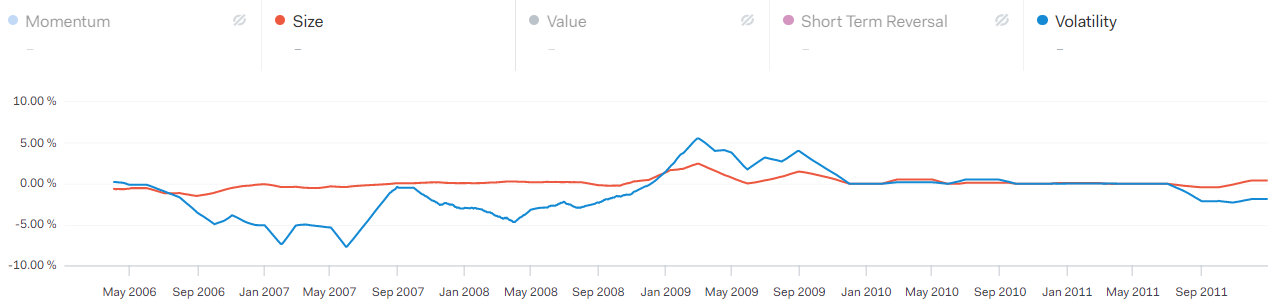}
    \caption{Size and Volatility Comparison for Kalman Filter Methodology}
    \label{fig 49: Size and volatility comparison for Kalman Filter Methodology}
\end{figure}

\begin{figure}[!htb]
    \centering
    \includegraphics[width=17cm, height=5cm]{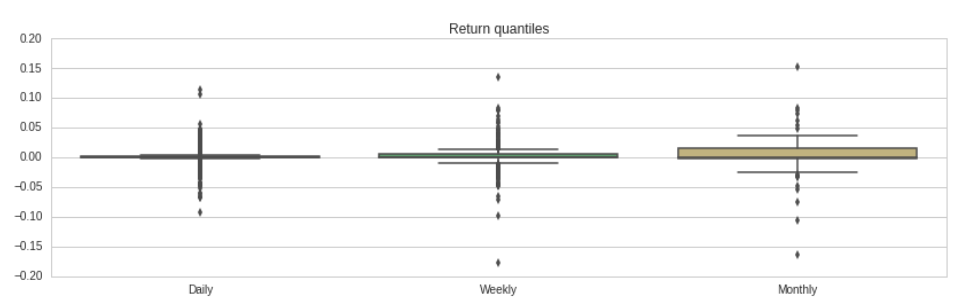}
    \caption{Return Quantiles over Day, Month, Week for Kalman Filter Methodology}
    \label{fig 50: Return Quantiles over Day, Month, Week for Kalman Filter Methodology}
\end{figure}

\clearpage

\begin{figure}[!htb]
    \centering
    \includegraphics[width=17cm, height=5cm]{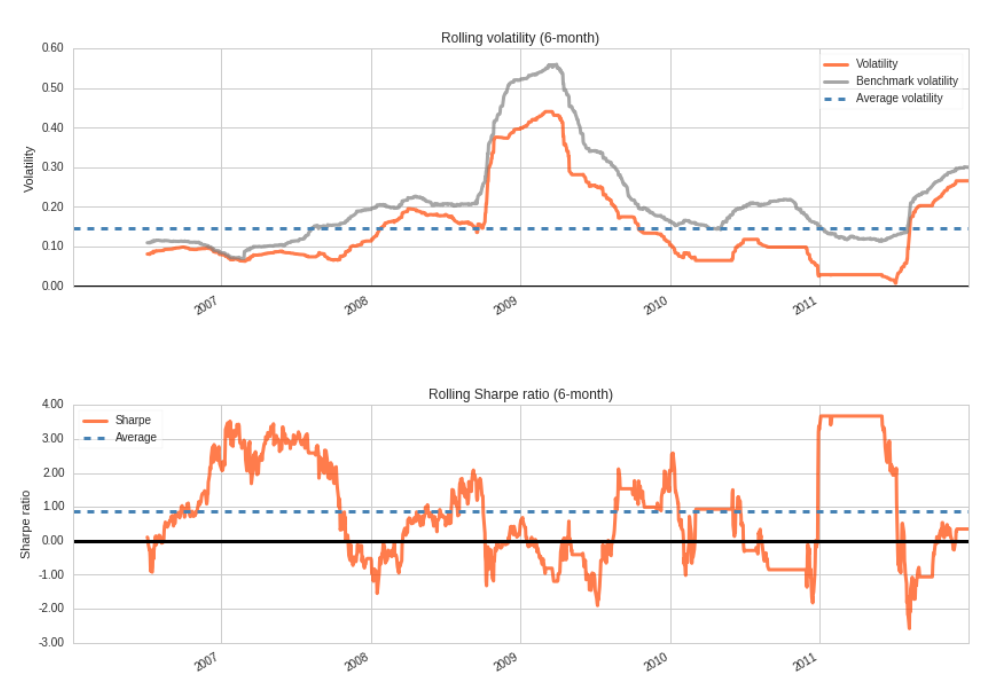}
    \caption{Rolling Volatility for Kalman Filter Methodology}
    \label{fig 51: Rolling Volatility for Kalman Filter Methodology}
\end{figure}

\begin{figure}[!htb]
    \centering
    \includegraphics[width=17cm, height=5cm]{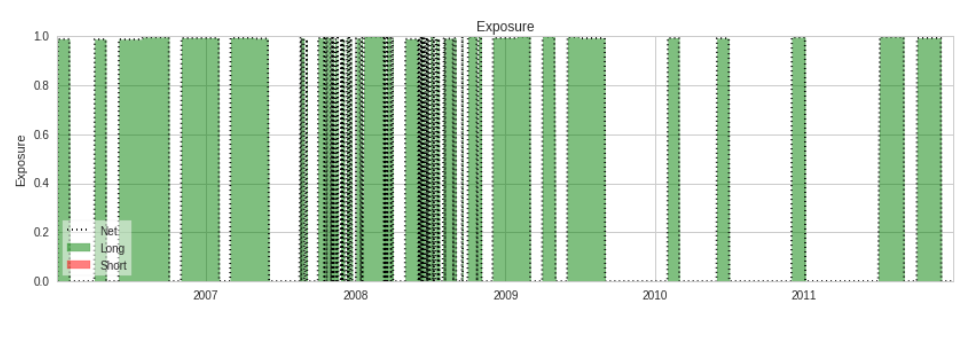}
    \caption{Exposure for long and short sizes for Kalman Filter Methodology}
    \label{fig 52: Exposure for long and short sizes for Kalman Filter Methodology}
\end{figure}

\begin{figure}[!htb]
    \centering
    \includegraphics[width=12cm, height=8cm]{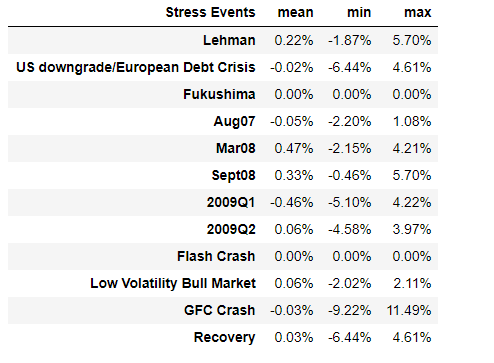}
    \caption{Performance of the backtest during high volatile situations for Kalman Filter Methodology}
    \label{fig 53: Performance of the backtest during high volatile situations for Kalman Filter Methodology}
\end{figure}

\clearpage

\subsection{Backtest using Minimum Variance}

\begin{figure}[h]
    \centering
    \includegraphics[width=4cm, height=3cm]{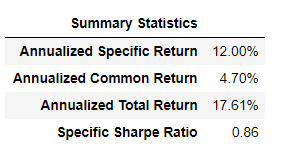}
    \caption{Performance Relative to Risk Factor}
    \label{fig 54: Performance Relative to Risk Factor Minimum Variance}
\end{figure}

\begin{figure}[!htb]
    \centering
    \includegraphics[width=8cm, height=9cm]{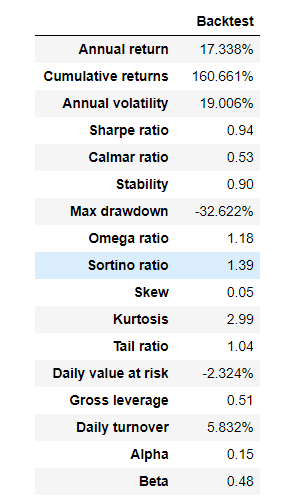}
    \caption{Backtest Information for Minimum Variance}
    \label{fig 55: Backtest Information for Minimum Variance}
\end{figure}

\clearpage

\begin{figure}[!htb]
    \centering
    \includegraphics[width=17cm, height=5cm]{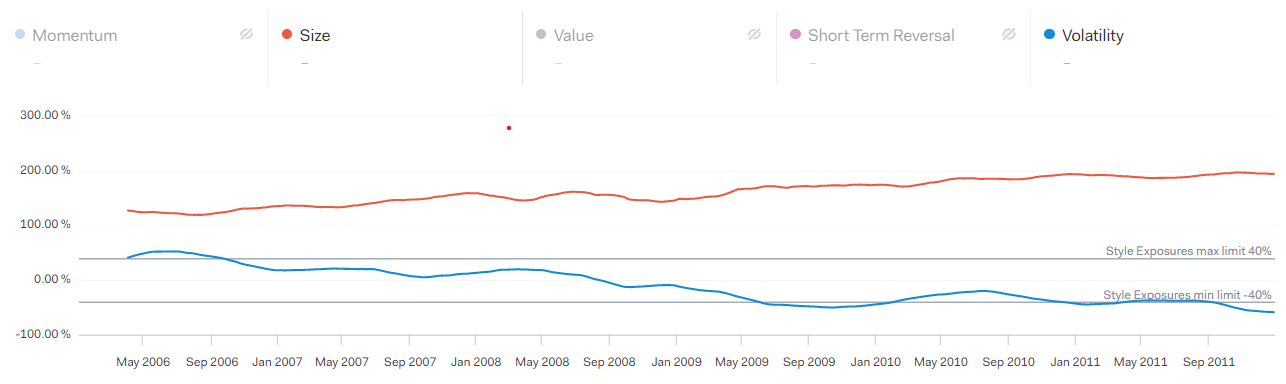}
    \caption{Size and Volatility Comparison for Minimum Variance}
    \label{fig 56: Size and volatility comparison for Minimum Variance}
\end{figure}

\begin{figure}[!htb]
    \centering
    \includegraphics[width=17cm, height=5cm]{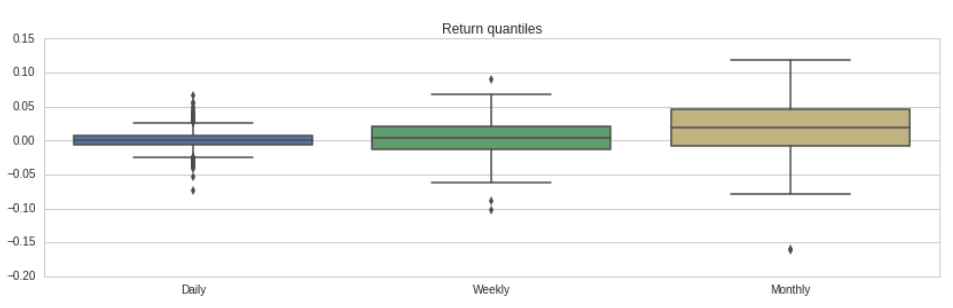}
    \caption{Return Quantiles over Day, Month, Week for Minimum Variance}
    \label{fig 57: Return Quantiles over Day, Month, Week for Minimum Variance}
\end{figure}

\clearpage

\begin{figure}[!htb]
    \centering
    \includegraphics[width=17cm, height=5cm]{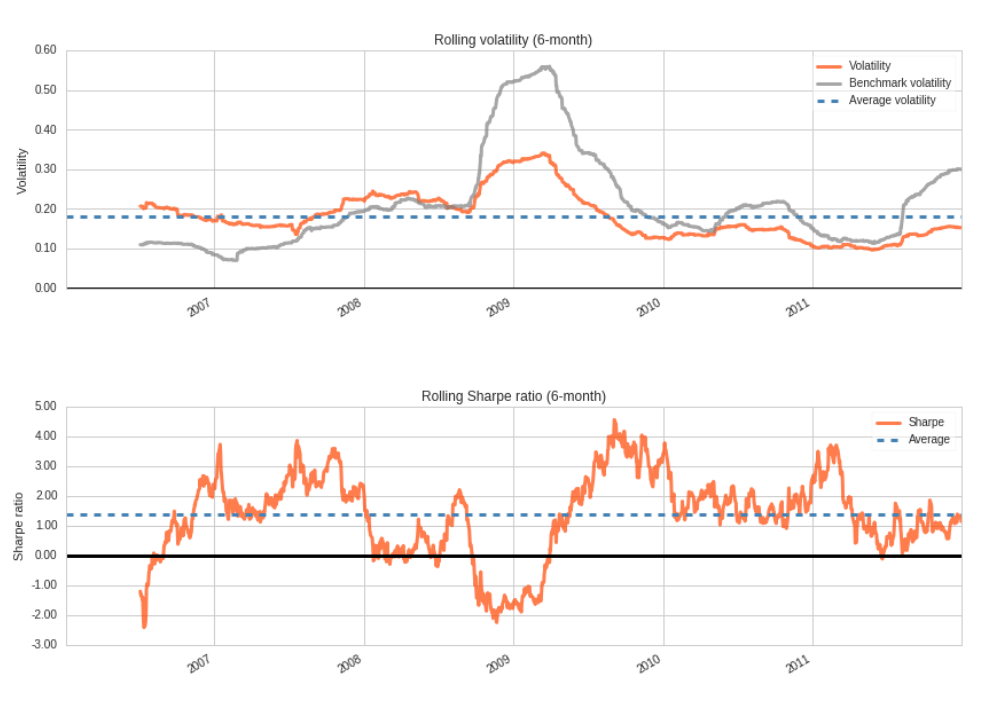}
    \caption{Rolling Volatility for Minimum Variance}
    \label{fig 58: Rolling Volatility for Minimum Variance}
\end{figure}

\begin{figure}[!htb]
    \centering
    \includegraphics[width=17cm, height=5cm]{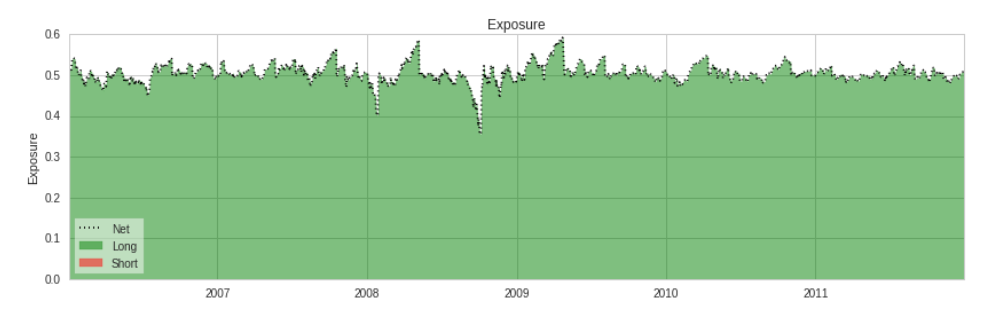}
    \caption{Exposure for long and short sizes for Minimum Variance}
    \label{fig 59: Exposure for long and short sizes for Minimum Variance}
\end{figure}

\begin{figure}[!htb]
    \centering
    \includegraphics[width=12cm, height=8cm]{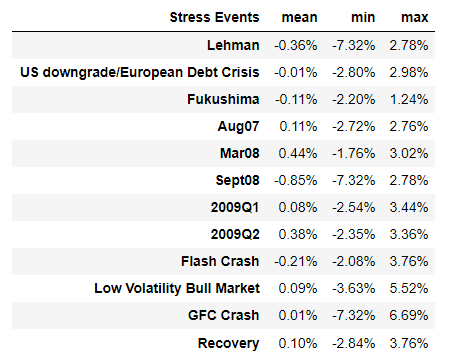}
    \caption{Performance of the backtest during high volatile situations for Minimum Variance}
    \label{fig 60: Performance of the backtest during high volatile situations for Minimum Variance}
\end{figure}

\chapter{Conclusion}
\section{Backtesting Results}
The backtesting results provide valuable insights into the effectiveness of various strategies, particularly within the context of Independence and Joint Testing, which comprised three distinct components:

\begin{enumerate}
    \item \textbf{Equal Short and Long Positions (Hedge):} \\
    The first part of this analysis involved maintaining an equal balance between short and long positions.

    \item \textbf{Long Positions Exceeding Short Positions:} \\
    In the second scenario, the long positions outweighed the short positions.

    \item \textbf{Short Positions Exceeding Long Positions:} \\
    The third scenario focused on situations where short positions were more substantial than long positions.
\end{enumerate}

Analyzing these scenarios in terms of performance reveals noteworthy findings. During periods of high volatility, it becomes evident that shorting stocks tends to yield higher returns. Additionally, long positions exhibit a positive impact during high volatility and Value at Risk (VaR) analysis. However, the concept of hedging, particularly during turbulent market conditions, may not be a prudent strategy.A particularly intriguing outcome emerged from the analysis using the Kalman filter methodology. This approach did not perform well when confronted with high levels of volatility. The Kalman filter method appeared to rely heavily on long positions, and it struggled to meet predefined parameters, indicating that it may not be well-suited for turbulent market conditions.In contrast, among all the backtesting strategies employed, the Variance test demonstrated the most promising results in terms of returns. It displayed resilience even in high-volatility scenarios. However, it's worth noting that the Variance test exhibited higher volatility compared to other methods. This strategy predominantly depended on long positions.\\

In conclusion, the results of these backtesting exercises collectively affirm the efficacy of a sizing model in mitigating risks during periods of high market volatility. Such a model can effectively reduce Value at Risk (VaR) and potentially offer a more stable approach to investment in challenging market conditions.

\phantomsection
\printbibliography
\addcontentsline{toc}{chapter}{Bibliography}

\end{document}